# ANALYSIS OF BIOMEDICAL SIGNALS BY FLICKER-NOISE SPECTROSCOPY: IDENTIFICATION OF PHOTOSENSITIVE EPILEPSY USING MAGNETOENCEPHALOGRAMS

[1,2] S.F. Timashev, [*2] Yu.S. Polyakov, [3,4] R.M. Yulmetyev, [3,4] S.A. Demin, [3,4] O.Yu. Panischev, [5] S. Shimojo, [6,7] J. Bhattacharya

[1] Karpov Institute of Physical Chemistry, ul. Vorontsovo pole 10, Moscow 105064, Russia

[2] USPolyResearch, Ashland, PA 17921, USA

[3] Department of Physics, Kazan State University, ul. Kremlevskaya 18, Kazan 420008, Russia

[4] Department of Physics, Kazan State Pedagogical University, ul. Tatarstan 2, Kazan 420021, Russia

[5] Division of Biology, California Institute of Technology, MC 139-74, Pasadena, CA 91125, USA

[6] Commission for Scientific Visualization, Austrian Academy of Sciences, Tech Gate, Vienna A 1220, Austria

[7] Department of Psychology, Goldsmiths College, University of London, New Cross, London SE146NW, UK

## ABSTRACT

The flicker-noise spectroscopy (FNS) approach is used to determine the dynamic characteristics of neuromagnetic responses by analyzing the magnetoencephalographic (MEG) signals recorded as the response of a group of control human subjects and a patient with photosensitive epilepsy (PSE) to equiluminant flickering stimuli of different color combinations. Parameters characterizing the analyzed stochastic biomedical signals for different frequency bands are identified. It is shown that the classification of the parameters of analyzed MEG responses with respect to different frequency bands makes it possible to separate the contribution of the chaotic component from the overall complex dynamics of the signals. It is demonstrated that the chaotic component can be adequately described by the anomalous diffusion approximation in the case of control subjects. On the other hand, the chaotic component for the patient is characterized by a large number of high-frequency resonances. This implies that healthy organisms can suppress the perturbations brought about by the flickering stimuli and reorganize themselves. The organisms affected by photosensitive epilepsy no longer have this ability. This result also gives a way to simulate the separate stages of the brain cortex activity *in vivo*. The examples illustrating the use of the "FNS device" for identifying even the slightest individual differences in the activity of human brains using their responses to external standard stimuli show a unique possibility to develop the "individual medicine" of the future.

---

[*] Corresponding author: ypolyakov@uspolyresearch.com

PACs: 02.70.Hm, 87.50.W-, 87.85.dm, 87.85.Ng, 89.75.-k

Topic: Laser Methods in Chemistry, Biology, and Medicine

1. INTRODUCTION

Many studies into the dynamics of living organisms and their subsystems, including the analysis of electroencephalograms, electrocardiograms, and tremor velocities of Parkinsonian patients [1-9], demonstrate that the real time series $V(t)$, where $t$ is the time, for measured dynamic variables, which characterize the current condition of a living system, on some specific time intervals $T$ usually contain chaotic components with a "long memory". In other words, there are correlation links on large intervals $T_{tot}$ of the time series. The links are commonly determined by analyzing the power spectrum $S_P(f)$, where $f$ is the frequency, when $S_P(f)$ is described by the flicker-noise function $f^{-n}$ with exponent $n \sim 1$ [6, 7]. In some cases, special "memory functions" characterizing the long-range correlations are introduced [8, 9]. The long-range links virtually indicate that the living organism, which is an open nonstationary system operating under variable external conditions, can quickly and efficiently reorganize itself, thus manifesting its property of biological adaptability [8]. Specifically, the analysis of heartbeat interval series shows that the mechanisms of neurohormonal cardiac regulation are brought into effect as dynamical rearrangements in the visual chaos of the studied time series. The decrease in chaoticity and the loss of long-range correlations in the measured biomedical signals may sometimes be associated with unfavorable changes in the organism, deviation of its functioning state from the normal state, and pathological changes in the organs [2-5].

As the real signals generated by living systems contain both chaotic and regular components [1, 8, 9], the above conceptual conclusions may not be directly used for evaluating the state of a specific organism and the effect of various factors such as medications and stimulation on its operation. In this case, it is necessary to separate the contribution of each component from the informative medical characteristics of the system before making any conclusions about the degree of loss in the correlation links. For example, stochastic quantifiers of a statistical memory were used to describe the phenomenological regularity accounting for the fundamental role of chaoticity and robustness in the functioning of living systems [8]. Based on the fundamental laws and concepts of memory functions formalism and FNS phenomenology, an original algorithm in which the effects of dynamic intermittency, nonstationarity, and characteristic frequencies in the original time series of a pathological tremor are separated out was proposed to evaluate the effectiveness and quality of different methods of treating Parkinson's disease [9]. This separation procedure can be used to advance in solving the general problems of medical diagnosis by providing the basis for the standards of medical signals through relating the values of specific signal characteristics to the particular state of the organism [10]. It is obvious that the general analysis of such problems is complicated by the presence of individual features in each organism and the specificity of its response to various treatment techniques.

The individual features of a living organism manifest themselves primarily in the low-frequency components of its biomedical signals, which account for the collection of characteristic and stimulus-initiated frequencies particular to each organism, as well as in the interferential contributions of these resonances. In this case, the low-frequency "envelopes" are always accompanied by high-frequency chaotic ("noise") components the series of which is used to identify the informative correlation links individual to each organism. The state of the living organism exposed to external, including therapeutic, stimuli and the dynamics of its subsystems can be adequately evaluated only by sequential separation of the contributions of these components in the physiological time series for different frequency bands and introduction of the corresponding parameterization.

The phenomenological scheme for representing the information stored in various complex signals developed by Nicholis [11] provided a conceptual foundation for making some progress in the practical extraction of both chaotic and resonant components from medical time series. This scheme assumes that there is an infinite number of levels in the evolution hierarchy of the system under study and that there are some recurrent rules that generate information on a specific hierarchy level and compress it on a higher level. The general ideas of Nicholis' scheme were used to develop flicker-noise spectroscopy (FNS) [12, 13], the phenomenological framework in which the concept of the information contained in the signals generated by open dissipative systems is generalized. According to the basic idea of FNS, the correlation links existing in the sequences of different irregularities such as spikes and "jumps" in original signals and discontinuities in their derivatives of different orders on all levels of the spatiotemporal hierarchy of the system under study are treated as main information carriers. In this theory, the tools for the extraction and analysis of information are the cosine-transform $S(f)$ of an autocorrelation function, where $f$ is the frequency, and the difference moment, or structural function, $\Phi^{(2)}(\tau)$ of the second order, where $\tau$ is the time lag, which complement each other in the extraction and analysis. The chaotic component $\Phi_c^{(2)}(\tau)$ of the structural function is formed exclusively by the "jumps" of the dynamic variable while the chaotic component $S_c(f)$ of the cosine-transform is formed by both spikes and "jumps" on every level of the hierarchy. The information "passport" characteristics determined by fitting the derived expressions to the experimental curves of $S_c(f)$ and $\Phi_c^{(2)}(\tau)$ are interpreted as the correlation times and parameters describing the rate of "memory loss" on these correlation time intervals for different irregularities. The number of extracted parameters depends on the problem under study [12-15].

FNS can be applied to three types of problems: (1) Identification of the parameters that characterize the dynamics or structural features of open complex systems. (2) Identification of the precursors of abrupt changes in the state of various open dissipative systems using the *a priori* information about the dynamics of the systems. (3) Determination of the flow dynamics in distributed systems by analyzing the dynamic correlations in the chaotic signals that are simultaneously measured at different points in space. The FNS approach has already been applied to the analysis of the structure and dynamics for various physicochemical, electrochemical, biological, geophysical, and astrophysical processes [12-15].

The major problem emerging in the analysis of biomedical signals is its ability to provide an adequate evaluation of the dynamic states of the organism. As the spatiotemporal structure of the signals of a living system generated as the response to given stimuli is unique, the adequacy of the analysis can be provided only when the method in use makes it possible to identify even the slightest individual differences in the measured signals. In the present paper, the study of neuromagnetic responses by analyzing the magnetoencephalographic (MEG) signals recorded as the response of a group of control subjects and a PSE patient to equiluminant flickering stimuli of different color combinations [16-19] will show that the FNS method can be used to identify even the slightest differences in the recorded individual responses.

## 2. BASIC PRINCIPLES OF FLICKER-NOISE SPECTROSCOPY

(1) A hierarchy of spatiotemporal levels in complex open dissipative systems whose chaotic evolution is described by measured dynamic variable $V(t)$ on time interval $T$ is introduced.

(2) The main information hidden in stochastic signals $V(t)$ is provided by specific "resonant" components as well as by sequences of different types of irregularities such as spikes and jumps in the original signals and discontinuities in their derivatives of different orders on all levels of the spatiotemporal hierarchy of the system under study. In FNS, all the introduced information is related to one of the fundamental concepts of statistical physics, the autocorrelation function

$$\psi(\tau) = \langle V(t)V(t+\tau) \rangle, \qquad (1)$$

where $\tau$ is the time lag parameter ($0 \leq \tau \leq T/2$). This function characterizes the correlation of the values of dynamic variable $V(t)$ at higher and lower values of the argument. The angular brackets in relation (1) stand for the averaging over time interval $T$:

$$\langle (...) \rangle = \frac{1}{T} \int_{-T/2}^{T/2} (...) dt. \qquad (2)$$

The averaging over interval $T$ implies that all the characteristics that can be extracted by analyzing $\psi(\tau)$ functions should be regarded as the average values on this interval. If the interval $T$ is a section of the larger interval $T_{tot}$ ($T < T_{tot}$), the value of function $\psi(\tau)$ can depend on the position of the interval $T$ within the larger interval $T_{tot}$. If there is no such dependence and $\psi(\tau)$ is a function only of the difference of the arguments of the dynamic variables involved in (2), the evolution process being analyzed is defined as stationary. In this case, $\psi(\tau) = \psi(-\tau)$.

(3) To extract the information contained in $\psi(\tau)$ ($\langle V(t) \rangle = 0$ is assumed), one should analyze the transforms, or "projections", of this function, specifically the cosine transform (power spectrum function) $S(f)$, where $f$ is the frequency:

$$S(f) = \int_{-T/2}^{T/2} \langle V(t)V(t+t_1) \rangle \cos(2\pi f t_1) dt_1 \qquad (3)$$

and its difference moments (transient structural functions) of the second order $\Phi^{(2)}(\tau)$:

$$\Phi^{(2)}(\tau) = \langle |V(t) - V(t+\tau)|^2 \rangle. \qquad (4)$$

It is obvious that for the stationary process we have

$$\Phi^{(2)}(\tau) = 2[\psi(0) - \psi(\tau)], \qquad (5)$$

implying that $\Phi^{(2)}(\tau)$ depends linearly on $\psi(\tau)$. The introduction of $\Phi^{(2)}(\tau)$ as the "projection" of autocorrelator $\psi(\tau)$ is convenient because the functions $\Phi^{(2)}(\tau)$ are formed solely by the jumps of the dynamic variable at different spatiotemporal hierarchy levels of the system and $S(f)$ is formed by both spikes and jumps.

To illustrate this statement, consider the process of one-dimensional "random walk" with small "kinematic viscosity" $v$ (Fig. 1). The small value of $v$ implies that when the signal passes from position $V_i$ to $V_{i+1}$, which are $|V_{i+1} - V_i|$ apart (in value) from each other, the system first overleaps ("overreacts") due to inertia and then "relaxes". We assume that the relaxation time is small compared to the residence time in the "fluctuation" position. It is obvious that when the number of walks is large, the functions $\Phi^{(2)}(\tau)$ will be independent of the values of "inertial overleaps" of the system and depend only on the algebraic sum of walk "jumps". At the same time, the functions $S(f)$, which characterize the "energy side" of the process, will depend on both spikes and jumps. It should be emphasized that it is the intermittent character of the evolution dynamics that accounts for the differences in the information stored in various irregularities. In other words, if there is no intermittency, the information contents of $S_c(f)$ and $\Phi_c^{(2)}(\tau)$ will coincide with each other.

(4) The "passport" parameters extracted by analyzing the functions $S(f)$ and $\Phi^{(2)}(\tau)$ built using the time series $V(t)$ have the meaning of relaxation times and the parameters characterizing the loss of "memory" (correlation links) at these correlation times for irregularities such as spikes and jumps. The corresponding parameters for irregularities such as "discontinuities in derivatives" are extracted from the power spectra and difference moments built using time series $\Delta^m V(t_k)/\Delta t^m$ ($m \geq 1$). Here, $\Delta^m V(t_k) = \Delta^{m-1} V(t_k) - \Delta^{m-1} V(t_{k-1})$; $\Delta t = t_k - t_{k1}$ is the sampling interval for the values of the dynamic variable recorded at discrete times $t_k$.

(5) Stationary processes in open dissipative systems, when the autocorrelator $\psi(\tau) = <V(t)V(t+\tau)>$ depends only on the difference of arguments $\tau$, are characterized by a multi-parameter self-similarity, in contrast to the single-parameter self-similarity in the fractal and renormgroup theories, which implies that each introduced parameter has the same value for every spatiotemporal hierarchy levels of the system.

(6) Expressions (3) and (4), as well as the interpolation relations for $S(f)$ and $\Phi^{(2)}(\tau)$ given below, are introduced by considering the function $V(t)$ determined at every point, which formally corresponds to the signal recorded at an "infinite" sampling frequency. At the same time, the raw data on the dynamics of real complex systems are usually obtained as the discrete time series $V(t_k)$ recorded at some finite sampling frequency $f_d$, where $t_k$ are the times separated by a fixed interval $\Delta t = f_d^{-1}$. When information is extracted from these discrete time series, first we should answer the fundamental question: How complete and reliable is the information contained in the signals recorded at some finite sampling frequency in view of the fact that real systems also generate signals at much higher frequencies? The Nyquist–Shannon–Kotelnikov theorem implies that in order to obtain reliable information about a resonant (regular) component with frequency $f_r$, the inequality $f_d \geq 2 f_r$ must be true [20]. In the frequency band from $1/T$ to $f_d/2$, where $f_d$ is the sampling frequency, the resonant components contribute mostly to the low-frequency band while all the irregularities manifest themselves mostly in the high-frequency bands, which include a high-frequency spike (high-frequency chaotic) band and a high-frequency jump (low-frequency chaotic) band. The change in the sampling interval causes changes in the introduced information parameters. The differences in the information contents of functions $S(f)$ and $\Phi^{(2)}(\tau)$, which are obvious for continuous signals (Fig. 1) and caused by the difference in the contributions made by the high-frequency (spike) and low-frequency (jump) chaotic components into these functions, can also manifest themselves in the functions $S(f)$ and $\Phi^{(2)}(\tau)$ build using discretely recorded time series [15]. This supports the validity of the fundamental FNS hypothesis that the difference in the information character of functions $S(f)$ and $\Phi^{(2)}(\tau)$ can exist even when there are no "singularities" in the measured discrete signals. In the general case, when a complex stochastic signal is measured at some sampling frequency $f_d$, we can speak only about finding the collection of parameters characterizing the correlation links in the sequences of jumps and spikes inherent to the given signal determined at sampling frequency $f_d$. Consequently, the sampling interval should be considered as another parameter in general parameterization algorithms for real signals.

## 3. BASIC RELATIONS

Let us write the basic interpolation expressions for chaotic components used in the analysis of experimental time series. The parameters characterizing the dynamic correlations on every level of the evolution hierarchy are assumed to be the same. Consider the simplest case, in which there is only one characteristic scale in the sequences of spikes and jumps:

$$\Phi_c^{(2)}(\tau) \approx 2\sigma^2 \cdot \left[1 - \Gamma^{-1}(H_1) \cdot \Gamma(H_1, \tau/T_1)\right]^2, \qquad (6)$$

$$\Gamma(s,x) = \int_x^\infty \exp(-t) \cdot t^{s-1} dt, \quad \Gamma(s) = \Gamma(s,0),$$

where $\Gamma(s)$ and $\Gamma(s, x)$ are the complete and incomplete gamma functions ($x \geq 0$ and $s > 0$), respectively; $\sigma$ is the standard deviation of the measured dynamic variable with dimension $[V]$; $H_1$ is the Hurst constant, which describes the rate at which the dynamic variable "forgets" its values on the time intervals that are less than the correlation time $T_1$ [12-15]. In this case, $T_1$ may be interpreted as the correlation time for the jumps in the stochastically varying time series $V(t)$.

For asymptotic cases, we obtain the formulas:

$$\Phi_c^{(2)}(\tau) = 2\Gamma^{-2}(1+H_1) \cdot \sigma^2 \left(\frac{\tau}{T_1}\right)^{2H_1}, \quad \text{if } \frac{\tau}{T_1} \ll 1 \qquad (7)$$

$$\Phi_c^{(2)}(\tau) = 2\sigma^2 \left[1 - \Gamma^{-1}(H_1) \cdot \left(\frac{\tau}{T_1}\right)^{H_1-1} \exp\left(-\frac{\tau}{T_1}\right)\right]^2, \quad \text{if } \frac{\tau}{T_1} \gg 1. \qquad (8)$$

The interpolating function for power spectrum component $S_{cS}(f)$ formed by spikes can be written as:

$$S_{cS}(f) \approx \frac{S_{cS}(0)}{1 + (2\pi f T_0)^{n_0}} \qquad (9)$$

Here, $S_{cS}(0)$ is the parameter characterizing the low-frequency limit of $S_{cS}(f)$ and $n_0$ describes the degree of correlation loss in the sequence of spikes on the time interval $T_0$.

The interpolating function for the power spectrum component $S_{cR}(f)$ formed by jumps is written as:

$$S_{cJ}(f) \approx \frac{S_{cJ}(0)}{1 + (2\pi f T_1)^{2H_1+1}}, \qquad (10)$$

where $S_{cJ}(0)$ is the parameter characterizing the low-frequency limit of $S_{cJ}(f)$.

Although the contributions to the overall power spectrum $S_c(f)$ given by Eqs. (9) and (10) are similar, the parameters in these equations are different: $S_{cS}(0) \neq S_{cR}(0)$, $T_1 \neq T_0$, and $2H_1 + 1 \neq n_0$. This implies that the parameters in the expressions for the power spectrum and structural function of the second order generally have different information contents when the experimental time series $V(t)$ is analyzed. For example, the characteristic times $T_0$ and $T_1$ are usually much different because they correspond to different frequency bands of the power spectrum. The fact that the jumps are more regular than the spikes implies that the contribution of jumps to the power spectrum will be concentrated in its lower frequency band. At the same time, its higher frequency band, which is often characterized by the flicker-noise function $S_c(f) \sim 1/f^n$, is generated mostly by spikes.

As noted above, the slowly varying components specific to each complex system under study, which are characterized by their own collection of frequencies, can affect the chaotic dynamics of the sequences of informative irregularities in such systems. This gives rise to interferential frequencies, which are produced by the addition of different resonant components, in the power spectra $S(f)$ to be analyzed. During the evolution of open systems, the entire set of these resonant and interferential frequencies can be rearranged. For convenience, in our further consideration all specific frequencies manifesting themselves in the oscillations of the dynamic variable $V(t)$ under study, no matter the genesis of these frequencies used in functions $S(f)$, will be considered as resonant.

## 4. SEPARATION OF FREQUENCY COMPONENTS FROM A SIGNAL

In the general case, the signal $V(t)$ under study can be formally written as

$$V(t) = V_{cS}(t) + [V(t) - V_{cS}(t)] = V_{cS}(t) + V_{cJ}(t) + V_r(t), \tag{11}$$

where

$$V_r(t) \equiv V(t) - V_{cS}(t) - V_{cJ}(t).$$

Here, $V_{cS}(t)$ and $V_{cJ}(t)$ are the chaotic components formed by spikes (mostly, the highest-frequency band) and jumps (mostly, the intermediate-frequency band), respectively; $V_r(t)$ is the low-frequency signal formed by resonant components, which are characterized by a "gradual" variation against the background of mostly high-frequency chaotic components.

Assume that there is no correlation of the high-frequency components $V_{cS}(t)$ with $V_r(t)$ and $V_{cJ}(t)$:

$$\begin{aligned}\psi_{rcS}(\tau) &\equiv \langle V_r(t)V_{cS}(t+\tau)\rangle = \langle V_{cS}(t)V_r(t+\tau)\rangle = 0, \\ \psi_{cSJ}(\tau) &\equiv \langle V_{cS}(t)V_{cJ}(t+\tau)\rangle = \langle V_{cJ}(t)V_{cS}(t+\tau)\rangle = 0\end{aligned} \tag{12}$$

and the later two components $V_{cJ}(t)$ and $V_r(t)$ can correlate with each other:

$$\psi_{rcJ}(\tau) \equiv \langle V_r(t)V_{cJ}(t+\tau) \rangle = \langle V_{cJ}(t)V_r(t+\tau) \rangle \neq 0. \tag{13}$$

In addition, it is assumed that

$$\psi_r(\tau) \equiv \langle V_r(t)V_r(t+\tau) \rangle = \psi(-\tau); \quad \psi_{rcJ}(\tau) = \psi_{rcJ}(-\tau);$$
$$\psi_{cS}(\tau) \equiv \langle V_{cS}(t)V_{cS}(t+\tau) \rangle \neq \psi_{cS}(-\tau); \quad \psi_{cJ}(\tau) \equiv \langle V_{cJ}(t)V_{cJ}(t+\tau) \rangle \neq \psi_{cJ}(-\tau). \tag{14}$$

Relations (12)-(14) make it possible to parameterize the signal $V(t)$ under study. First, it is necessary to subtract the interpolation function $S_{cS}(f)$ [expression (9)], which corresponds to the contribution of high-frequency components $V_{cS}(t)$ to $S(f)$, from the spectrum $S(f)$. Then, the use of the method of least squares allows us to determine the parameters $n_0$ and $T_0$ for each chosen value of $S_{cS}(0)$. In this case, the later value is optimized at the last stage of the parameterization procedure under consideration. Actually, the subtracted function $S_{cS}(f)$ can adequately approximate only the band of high frequency values accounting for the parameter $n_0$. Therefore, the determined value of $T_0$ should be corrected below. In this case, this value depends on the contribution made by the jumps in the band of intermediate frequencies.

The resulting difference $Q_{rcJ}(f) = S(f) - S_{cS}(f)$ in view of (3) and (12)-(14) can be written as

$$Q_{rcJ}(f) = S_r(f) + 2S_{rcJ}(f) + S_{cJ}(f), \tag{15}$$

where $S_r(f)$, $S_{cS}(f)$, and $S_{cS}(f)$ are the cosine transforms corresponding to the correlators $\psi_r(\tau)$, $\psi_{rcJ}(\tau)$, and $\psi_{cJ}(\tau)$. The function $Q_{rcJ}(f)$ can be used to write the "incomplete" inverse cosine transform:

$$\varphi_{rcJ}(\tau) = 2 \int_0^{f_{max}} Q_{rcJ}(f)\cos(2\pi f\tau)df, \tag{16}$$

where $f_{max} = \frac{1}{2} f_s = \frac{1}{2} \Delta t^{-1} = N/2T$; $f_s$ is the sampling frequency; $\Delta t$ is the time interval between the adjacent digitized values of signal $V(t)$, the total number of which is equal to $N$. It is obvious that even if conditions (12)-(14) are satisfied, there is no way to write expression (16) as a linear combination of introduced correlators $\psi_r(\tau)$, $\psi_{rcJ}(\tau)$ and $\psi_{cJ}(\tau)$ because of the bounded value of the upper integral limit in (16). Moreover, one can expect that the value of autocorrelator $\psi_{cJ}(\tau)$, which corresponds to the high-frequency (against the background of resonant frequency) component of the signal, will mostly contribute to $\varphi_{rcJ}(\tau)$ at small values of $\tau$ in the interval [0 ≤ $\tau_{max}$ ≤ T/2] under consideration. At the same time, the autocorrelator $\psi_r(\tau)$ and interferential cross-correlator $\psi_{rcJ}(\tau)$, which depend on the low-frequency resonant component, can be virtually reproduced [mainly, $\psi_r(\tau)$] using transform (16) and are able to determine the function $\varphi_{rcJ}(\tau)$ in the most part of interval $\tau$. The above is supported by the further analysis of signals $V(t)$ of different nature given in [12, 14] and below.

In the parameterization procedure, $\varphi_{rcJ}(\tau)$ in view of (5) and (14) is used to find the function:

$$\tilde{\Phi}_r^{(2)}(\tau) = 2\left[\varphi_{rcJ}(0) - \varphi_{rcJ}(\tau)\right], \tag{17}$$

which characterizes the total contribution made by the resonant component and interferential "resonant-jump" terms into the difference moment of the second order $\Phi^{(2)}(\tau)$. In this case, the complete function $\Phi^{(2)}(\tau)$ can be found by adding interpolation expression (6) to (17):

$$\Phi^{(2)}(\tau) = \Phi_c^{(2)}(\tau) + \tilde{\Phi}_r^{(2)}(\tau). \tag{18}$$

The analysis made in [12, 14] and below shows that the use of expression (18) for comparison with the complicated functions (4) based on the experimental values of time series $V(t)$ is quite reasonable. The comparison of experimental data and the values calculated by relations (17) and (18) using the method of least squares is used to determine the values of parameters $\sigma$, $T_1$, and $H_1$, which characterize the contribution made by jumps into the structural function. In this case, the values of parameters $H_1$, $\sigma$, and $T_1$ are chosen by providing the best agreement between the experimental and calculated curves $\Phi^{(2)}(\tau)$ in the entire interval of $\tau$ under study. As a result, the above parameter $H_1$ is somewhat different in meaning from the Hurst constant, which is usually introduced for describing the functions $\Phi^{(2)}(\tau)$ at small values of $\tau$. After that, the parameters $H_1$ and $T_1$ are used to find the interpolation function $S_{cJ}(f)$, formula (10), which characterizes the contribution made by jumps into the spectrum $S(f)$ of the autocorrelator. In this case, the value of $S_{cJ}(0)$, like the value of $S_{cS}(0)$ introduced above, is regarded as a free parameter.

When the interpolation spectrum found above,

$$S_{\text{int}}(f) = Q_{rcJ}(f) + S_{cS}(f) + S_{cJ}(f), \tag{19}$$

is compared with the overall spectrum $S(f)$, which was calculated by (3) using the experimental values of time series $V(t)$, with the help of the method of least squares, $T_0$ should be regarded as a free parameter as well, in addition to $S_{cJ}(0)$ and $S_{cS}(0)$. The value of this parameter should be corrected because of the obvious difficulty of adequately estimating the contribution of jumps in the intermediate frequency band of spectrum $S(f)$ at the first step of the parameterization procedure. If necessary, the parameters of resonances occurring in the process under study, such as the values of frequency $f_{0i}$, width $\gamma_i$, and intensity $A_i$, can be determined using the curve $Q_{rcJ}(f)$.

5. RELATION BETWEEN FNS AND DIFFUSION PARAMETERS

To conclude this section, we will discuss the relation between the introduced difference moments of the second order $\Phi^{(2)}(\tau)$, which in the general case characterize the random walks of system states (Fig. 1), and the processes of anomalous diffusion. For these processes, the average squared deviation of system states $V$ varying over the whole set of possible states, ($-\infty < V < \infty$), from the average value found for the time $\tau$ can be written as [21-25]

$$\langle (\Delta V)^2 \rangle_{pdf} = 2Dt_0 (\tau/t_0)^{2H_1}. \tag{20}$$

Here, $D$ is the diffusion coefficient; $t_0$ is the characteristic time; $H_1$ is the Hurst constant; the averaging, denoted by the symbol $\langle(…)\rangle_{pdf}$, is effected by introducing the probability density function $W(V, t)$, which accounts for the probability of the system state being within the given interval of states. It is assumed that the system was found in the vicinity of the $V = 0$ state (point) at the initial time $\tau = 0$. Fick's diffusion ($H_1 = ½$) corresponds to the random walks of system states characterized by some characteristic scale $\delta V$ of the values of elementary jumps when the system transfers between adjacent states, and by the characteristic residence time $\delta\tau$ for every state. However, if these random walks of states stochastically alternate with the jumps having anomalous values higher than $\delta V$ at the same characteristic residence times $\delta\tau$ for the given state, the so-called superdiffusion (Levy diffusion or flights), for which $H_1 > ½$, can occur. If the random walks stochastically alternate with the jumps having anomalously long times of residence in some states ("stability islands" [24]), which are much larger than $\delta\tau$, for the same values of jumps $\delta V$, the so-called subdiffusion, for which $H_1 < ½$, can occur.

Anomalous (from Fick's viewpoint) diffusion processes can be described using diffusion equations with constant diffusion coefficients in which the partial derivatives with respect to time and coordinate are replaced by fractional-order derivatives [21]. In this case, the subdiffusion process is described by introducing the fractional derivative of order $\alpha$ ($0 < \alpha < 1$) instead of the partial derivative of the first order in time whereas the superdiffusion process is described by introducing the fractional derivative of order $\beta$ ($1 < \beta < 2$) instead of the partial derivative of the second order with respect to coordinate [21-24]. The parameter $H_1$ is varied in the ranges $0 < H_1 < ½$ for subdiffusion and $½ < H_1 < 1$ for superdiffusion. It should be noted that if the process under study is more complicated, such as the one in which the diffusion coefficient depends on coordinate, the value of the Hurst constant can be higher than unity.

FNS is a phenomenological approach. The parameters introduced in FNS have a certain physical meaning and are determined by comparing the results calculated by FNS relations (5) – (10) with the curves found with (3) and (4) using the experimental values of $V(t)$ forming the time series. For a stationary process, in which the autocorrelator $\psi(\tau) = \langle V(t)V(t+\tau) \rangle$ depends only on the difference of the arguments of dynamic variables and it is assumed that the ergodicity condition is met, the procedure of averaging over time (2) introduced in FNS is equivalent to the averaging procedure using the probability density function $W(V, t)$. In this case, expression (4) can be regarded as the generalized expression for the average squared deviation from the average value in the random walk processes described by Fick's equation or the equations of anomalous diffusion.

To find the relation between the phenomenological FNS parameters $\sigma$, $T_1$ and $H_1$ and the parameters characterizing the diffusion dynamics for a stationary process, we will first consider the simplest case of Fick's diffusion, for which $H_1 = \frac{1}{2}$. Assume that the behavior of the probability density $W(V, \tau)$ for the random variable $V$ on the segment $[-L, +L]$ over time $\tau$ can be described by the diffusion equation:

$$\frac{\partial W}{\partial \tau} = D \frac{\partial^2 W}{\partial V^2} \tag{21}$$

with the reflection conditions at the end points of the segment

$$\frac{\partial W}{\partial V} = 0 \text{ at } V = -L \text{ and } V = +L \tag{22}$$

and the initial condition

$$W(V, 0) = \delta(V). \tag{23}$$

The expressions for the solution $W(V, \tau)$ to this system, the average value of random variable $V$, and the average squared deviation of this variable from its average value over time $\tau$ are given in Appendix: (A.1), (A.2), and (A.3), respectively. The relation between the parameters of the diffusion problem and phenomenological FNS parameters can be found by comparing asymptotic expressions (A.4) and (A.5) for small and large values of $\tau$ with corresponding expressions (7) and (8) written with $H_1 = \frac{1}{2}$:

$$D = \frac{4}{\pi} \cdot \frac{\sigma^2}{T_1}; \quad L^2 = 6\sigma^2. \tag{24}$$

In this case, the difference between the values calculated by expressions (A.3) and (4) with $H_1 = \frac{1}{2}$ for the range of intermediate values of the parameter $\tau$ does not exceed 20% (Fig. 2). The figure demonstrates the normalized curves calculated by (4) and (A.3), for which the asymptotic values at $x \ll 1$ and $x \gg 1$ coincide:

$$\varphi_1(x) \equiv \frac{1}{2\sigma^2} \Phi^{(2)}(T_1 x), \quad \varphi_2(x) \equiv \frac{3}{L^2} \langle V^2 \rangle_{pdf}, \tag{25}$$

where $x = \tau / T_1$.

It is difficult to compare the functions $\langle V^2 \rangle_{pdf}$ and $\Phi^{(2)}(\tau)$ for anomalous diffusion ($H_1 \neq \frac{1}{2}$) because the corresponding equations for the probability density $W(V, \tau)$ of random variable $V$ varying on the segment $[-L, +L]$ over the time $\tau$ are very complicated and can be solved only by numerical methods [24, 26]. At the same time, the desired relation can be found if we compare expressions (7) and (20), which correspond to the case of anomalous diffusion at small values of $\tau$, by choosing $T_1$ as the characteristic time $t_0$ and assume that in this case the FNS parameter $\sigma$ corresponds to some model parameter $L_a$ determining the region in the domain of values of the dynamic variable to be analyzed where the states of the system can be localized:

$$D = \frac{1}{\Gamma^2(1+H_1)} \cdot \frac{\sigma^2}{T_1}; \quad L_a^2 = b\sigma^2, \tag{26}$$

Here, $b$ is the dimensionless parameter. Below, biomedical signals will be analyzed to show how the concept of anomalous diffusion can be used to identify the individual functioning features of a human brain.

## 6. FNS ANALYSIS OF MAGNETOENCEPHALOGRAMS

In this section, the FNS approach will be used for comparative analysis of the dynamic characteristics of the neuromagnetic responses, such as magnetoencephalograms (MEGs), generated by the cerebral cortices of a 12-year-old patient with photosensitive epilepsy (PSE) and two healthy people in a control group, subjects 7 and 9. PSE is a common type of stimulus-induced epilepsy, defined as recurrent convulsions caused by visual stimuli, particularly by flickering light. The diagnosis of PSE involves finding paroxysmal spikes in an electroencephalogram in response to intermittent light stimulation. To find the color dependence of photosensitivity in normal subjects, they were exposed to uniform chromatic flickers and the whole-scalp MEGs of their brain activity were measured [16-17]. The analyzed responses were caused by flickering stimuli of different colors: red-blue (RB) and red-green (RG). The experimental setup shown in Fig. 3 displays the data generated by the 61-SQUID (superconducting quantum interference device) sensors attached to different points around the head, which can record weak magnetic induction gradients of about $10^{-11}$-$10^{-10}$ T/cm. The sampling frequency $f_d$ of MEG signals was 500 Hz ($f_d = 500\ Hz$). It was found that among the sensors recording MEG response signals, sensor 10, which is localized at the frontal lobe on the head, shows the highest sensitivity to these color stimuli [18, 19].

It was shown that the zones of the human cerebral cortex located in the vicinity of sensor 10, as well as some others, are the zones where the pathological PSE mechanisms are localized and can be the propagation centers for the abnormal group excitement of neurons in the cortex and subcortical structures, which ultimately causes an epileptic seizure [18, 19]. Therefore, the present paper was focused mostly on the analysis of the MEG response signal recorded by this sensor. Also, to compare the results, we analyzed the signal recorded by nonspecific sensor 43, which was located on the scalp over the occipital lobe of the cerebral cortex.

Figures 4, 5, and 6 illustrate the MEG signals recorded by sensor 10 as the response to the RB stimulus for healthy subjects 7, 9, and the PSE patient, respectively. The MEG signals recorded as the response to the RG stimulus are given in Figs. 1, 8, and 9, respectively. Every figure also demonstrates the results of FNS analysis of the recorded signals, with the FNS parameters being given in the figure captions.

First, we will compare the FNS parameters characterizing the responses of the healthy subjects and patient to the RB stimulus. The differences between the recorded responses of the subjects and patient primarily manifest themselves in the spectra $S(f)$. The low-frequency bands in the spectra of subjects 7 and 9 show a collection of specific peaks at frequencies from 2 to 30 Hz (Figs. 4b, 5b) whereas the patient spectrum shows clearly seen 50 and 100 Hz peaks in addition to the intense 2.5 Hz peak (Fig. 6b).

We can also see the difference in the characteristic curves for the components of $\tilde{\Phi}_r^{(2)}(\tau)$, which determine the total contribution of specific frequencies and the interferential "resonant-jump" component into the structural functions $\Phi^{(2)}(\tau)$. There are clearly cut differences not only in the values of the parameters, mainly in $T_1$, which are two orders of magnitude lower for the patient, but also in the presence of large intervals on the abscissa axis $\tau$ with $\tilde{\Phi}_r^{(2)}(\tau) < 0$ for either subject (Figs. 4c, 5c). At the same time, the corresponding curve for the patient response signal shows $\tilde{\Phi}_r^{(2)}(\tau) > 0$ throughout the entire range of $\tau$ (Fig. 6c). In this case, it is necessary to note the values of relaxation times $T_0$ found by analyzing the response signal produced by the patient, which are one order of magnitude shorter than those for the subjects.

A rather unexpected result of the analysis is the fact that the function typical for anomalous Levy diffusion can adequately approximate the chaotic (less the total contribution of resonances and interferential resonant-jump component) component in the signals produced by subjects 7 and 9 (Figs. 4d, 5d). In this case, the value of parameter $H_1$ for subject 9 is not much higher than the value of $H_1 = 0.5$, which is characteristic of Fick's diffusion. Consequently, this implies that even in a very complex process such as the cerebral neuroactivity we can separate out the components that can be described by physical and mathematical models. The construction of this chaotic component became possible only after the resonant components that continuously transform *in vivo* during the evolution and therefore defy mathematical simulation were "removed" from the analyzed functions. At the same time, the analysis of the patient signal showed that the function typical for anomalous diffusion cannot be used to describe the behavior of this chaotic component (Fig. 6d), which can be attributed mainly to the presence of a large number of high-frequency components in the original signal. Actually, anomalous diffusion has nothing to do with the dynamics of the patient signal, and the value of coefficient $D$ given in the caption to Fig. 4d, which is three orders of magnitude higher than the corresponding values for the signals of subjects 7 and 9, was used only for the formal comparison of the values of *D*. It is also necessary to note the anomalously high value of parameter $H_1 = 7.1$, which characterizes the nearly complete absence of correlations in the sequence of jumps in the signal (Poissonian process) recorded by sensor 10 in the patient as the response to the RB stimulus.

The data given in Figs. 7-9 show that the responses of the cerebral-cortex zone corresponding to sensor 10 to the RG and RB stimuli are not the same. Comparison of the parameters of the MEG responses of subjects 7 and 9 to the RB and RG stimuli shows that the MEG RG-response signals are characterized by the lower values of Hurst constants $H_1$ and effective diffusion coefficients *D* and the higher values of correlation times $T_1$ and $T_0$. This implies that the chaotic response signals caused by the RG stimulus are characterized by longer internal correlation links as compared to those caused by the RB stimulus.

The same behavioral pattern is observed for the parameters of the MEG response to the RG stimulus when we analyze the patient responses. The action of the RG stimulus leads to some decrease in the height of the 100 Hz high-frequency peak relative to the 50 Hz peak in the autocorrelator spectrum (Fig. 9b), as compared to the action of the RB stimulus, for which the 100 Hz peak is somewhat higher than the 50 Hz peak (Fig. 5b). As in the RB case, the RG-response chaotic component $\Phi^{(2)}(\tau)$ cannot be described by the Levy (anomalous diffusion) approximation (Fig. 9d). It should be noted that the value of component $\tilde{\Phi}_r^{(2)}(\tau)$ in the patient MEG signals recorded by sensor 10 is positive throughout the interval $\tau$ for both RG and RB stimuli (Fig. 9c).

Figures 10, 11, and 12 illustrate the MEG signals recorded by sensor 43 as the response to the RB stimulus for healthy subjects 7, 9, and the PSE patient, respectively. The MEG signals recorded as the response to the RG stimulus are shown in Figs. 13, 14, and 15, respectively. As in the analysis of the signals recorded by sensor 10, the differences between the responses recorded by sensor 43 for the subjects and patient manifest themselves in the spectra *S(f)*, though they are not so large as those for sensor 10. The low-frequency bands in the spectra for subjects 7 and 9 show specific peaks for frequencies up to 20-40 Hz (Figs. 10b, 11b, 13b, 14b) whereas the collection of these frequencies for the patient is complemented by a rather weak peak at a frequency of 50 Hz (Figs. 12b, 15b). The curves $\tilde{\Phi}_r^{(2)}(\tau)$ plotted for sensor 43, which determine the total contribution of specific frequencies and interferential resonant-jump components into the structural functions $\Phi^{(2)}(\tau)$, have large intervals on the abscissa axis $\tau$ in which the values of $\tilde{\Phi}_r^{(2)}(\tau)$ are negative not only for the signals recorded in subjects 7 and 9 in response to the RB and RG stimuli (Figs. 10c, 11c, 13c, 14c) but also for the patient signal recorded as the response

to the RG stimulus (Fig. 15c). In all these cases, the chaotic component of the structural function $\Phi^{(2)}(\tau)$ is quite adequately described by the approximation typical for anomalous diffusion (Figs. 10c, 11c, 13c, 14c, 15c). It should be noted that even for the RB stimulus, the curve $\tilde{\Phi}_r^{(2)}(\tau)$ plotted for patient's sensor 43 is negative in a small segment of the interval of $\tau$ (Fig. 12c). Although the corresponding chaotic component for the patient cannot be represented in the form typical for anomalous diffusion (Fig. 12c), the effective value of $D$ given in the figure caption may be used in making formal estimates.

The behavior of the parameters characterizing the structural function, such as $H_1$, $T_1$, and $T_0$, shows that their values for the MEG RB and RG responses recorded by sensor 43 in subjects 7 and 9 are not much different from the corresponding parameters determined by analyzing the patient signal recorded by sensor 43. It should be noted that each collection of these parameters is somewhat individual. For example, as in the analysis of the signal of sensor 10, the response of subject 7 to the RG stimulus is characterized by a higher internal correlation in the sequence of chaotic irregularities as compared with the response to the RB stimulus. At the same time, the MEG responses of subject 9 and the patient show a higher internal correlation in the "chaos" of signals recorded as the response to the RB stimulus. In this case, however, the entire set of reported data implies a relatively low diagnostic effectiveness of analyzing the signals produced by sensor 43 for PSE, which is in agreement with the previous general findings [18, 19].

Consequently, the FNS analysis of MEG signals recorded as the response to a color stimulus can be used to reliably identify the differences between the parameters of these signals for a patient and healthy subjects and quantify the symptoms of PSE, such as the anomalous group activity of cerebral cortex neurons, best identifiable by the analysis of the signals of sensor 10, which is accompanied by various clinical and paraclinical symptoms. These results are in agreement with the general findings of previous papers [18, 19]. For instance, the neuromagnetic responses of the cerebral cortex of a PSE patient show the obvious manifestation of additional quasi-periodic processes, which are absent in the signal of the magnetoelectrical activity of the brain of healthy people. Also, the above results are of interest to the problem of the regulating role of chaos in the dynamics of native systems, which has been discussed for a long time [2, 8]. Here, of importance is the behavior of found $\tilde{\Phi}_r^{(2)}(\tau)$ curves, which account for the total contribution of specific resonances and the interferential resonant-jump component into structural functions $\Phi^{(2)}(\tau)$. The important feature of these curves plotted for the MEG signals recorded by sensor 10 for both subjects as the responses to RB and RG stimuli is the presence of large negative intervals of these curves on the abscissa axis $\tau$. In contrast, the corresponding curve plotted using the patient MEG signals recorded by sensor 10 as the responses to RB and RG stimuli is positive throughout the studied interval $\tau$. In this case, the values of parameters $\sigma$, which characterizes the "measure" of fluctuations of the amplitude of the MEG response signal relative to the slowly varying resonant components, were rather close to each other for all considered cases. The latter means that if the contributions of $\tilde{\Phi}_r^{(2)}(\tau)$ into the structural functions $\Phi^{(2)}(\tau)$ characterizing the MEG responses of healthy subjects were positive, the *in vivo* values of average squared deviations would have been much higher.

This is not in conflict with the data obtained by the FNS analysis of the signals recorded by sensor 43, which is characterized by a low PSE diagnostic effectiveness. The contribution of components $\tilde{\Phi}_r^{(2)}(\tau)$ into the curve $\Phi^{(2)}(\tau)$ is negative on large intervals $\tau$ for the responses of both subjects to either type of stimulus and for the patient response to the RG stimulus. In the patient response to the RB stimulus, this component is negative only on a small segment of the interval $\tau$ and positive on the rest of it.

## 7. CONCLUSIONS

The above FNS analysis of MEG response signals demonstrates a principal possibility of separating the contribution of the diffusional component out of the overall complex dynamics of studied biomedical processes, implying that this can be used to simulate the separate functioning stages of a cerebral cortex *in vivo*.

The fact that the chaotic component of MEG signals can be adequately described by the anomalous diffusion approximation in the case of control subjects implies that healthy organisms can suppress the perturbations brought about by the flickering stimuli and reorganize themselves. The fact that the diffusion approximation cannot adequately describe the chaotic component for the patient may indicate that the organisms affected by photosensitive epilepsy lose this ability.

The demonstrated effectiveness of FNS analysis in identifying the individual features of the MEG responses of not only a patient but also healthy subjects to each of the RB and RG stimuli implies the possibility of using a FNS device for identifying even the slightest individual differences in the human cerebral activity in response to external standard stimuli. This allows us to effectively use the FNS methodology in developing the quantitative and qualitative methods of early diagnosis not only for photosensitive epilepsy but also for other neurodegenerative diseases, such as Parkinson's, Alzheimer's, Huntington's, amyotrophic lateral sclerosis, schizophrenia, and epilepsy, in order to identify and analyze the specific features of their course. The above data, as well as several other examples of FNS applications [12-15] demonstrating the principal possibility of identifying even the slightest individual differences in the responses of living systems, enable us to consider FNS as a rather promising methodology for the individual medicine of the future.

This study was supported in part by the Russian Foundation for Basic Research, projects no. 08-02-00230*a* and 08-02-00123*a*.

APPENDIX

Writing the Dirac delta function $\delta(V)$ as the series [27]:

$$\delta(V) = \frac{1}{2L}\left[1 + 2\sum_{k=1}^{\infty} \cos\frac{\pi k V}{L}\right],$$

we can find the solution to Eq. (21) subject to the above initial and boundary conditions:

$$W(V,\tau) = \frac{1}{2L}\left[1 + 2\sum_{k=1}^{\infty} \exp\left(-\frac{\pi^2 k^2 D\tau}{L^2}\right)\cos\frac{\pi k V}{L}\right]. \tag{A.1}$$

Now we can obtain the expressions for the average value of random variable $V$ and the average squared deviation of this variable from the average value in time $\tau$:

$$<V>_{pdf} = \frac{\int_{-L}^{+L} V W(V,\tau) dV}{\int_{-L}^{+L} W(V,\tau) dV} = 0, \tag{A.2}$$

$$\left\langle V^2 \right\rangle_{pdf} = \frac{\int_{-L}^{+L} V^2 W(V,\tau) dV}{\int_{-L}^{+L} W(V,\tau) dV} = \frac{4L^2}{\pi^2} \sum_{k=1}^{\infty} \frac{(-1)^{k+1}}{k^2} \left[ 1 - \exp\left(-\frac{\pi^2 k^2 D\tau}{L^2}\right) \right]. \tag{A.3}$$

From (24), we can derive the asymptotic expressions:

$$\left\langle V^2 \right\rangle_{pdf} \to 2D\tau \qquad \text{when } \tau \ll \frac{L^2}{\pi^2 D}, \tag{A.4}$$

$$\left\langle V^2 \right\rangle_{pdf} \to \frac{L^2}{3} \qquad \text{when } \tau \gg \frac{L^2}{\pi^2 D}. \tag{A.5}$$

Asymptotic expression (A.5) was found using the formula [28]:

$$\sum_{k=1}^{\infty} \frac{(-1)^{k+1}}{k^2} = \frac{\pi^2}{12},$$

Expression (A.1) was found using the first derivative of (A.3) with respect to time $\tau$:

$$\frac{d \left\langle V^2 \right\rangle_{pdf}}{d\tau} = 4D \sum_{k=1}^{\infty} (-1)^{k+1} \exp(-\xi k^2), \tag{A.6}$$

where $\xi \equiv \pi^2 D\tau/L^2 \ll 1$, in view of the relation:

$$\sum_{k=1}^{\infty} (-1)^{k+1} \exp(-\xi k^2) = \sum_{n=1}^{\infty} \Delta n \left\{ \exp\left[-4\xi\left(n-\frac{1}{2}\right)^2\right] - \exp(-4\xi n^2) \right\} =$$

$$= \frac{1}{2\xi} \left[ \int_0^{2\xi} \exp\left(-\frac{x^2}{\xi}\right) dx - \int_0^{\xi} \exp\left(-\frac{x^2}{\xi}\right) dx \right] = \frac{1}{4}\left(\frac{\pi}{\xi}\right)^{1/2} \left[ \Phi\left(2\xi^{1/2}\right) - \Phi\left(\xi^{1/2}\right) \right] \xrightarrow[\xi \to 0]{} \frac{1}{2}, \tag{A.7}$$

where $\Phi(x)$ is the error integral [28]. In (A.7), the change in the discrete values of $n$ in the summation was formally taken to be $\Delta n = 1$, followed by the transition from summation to integration using the integration variable $x = \xi n$ and the differential $d\xi = \xi$ when $\Delta n \ll 1$. Relation (26) was found by comparing (A.4) and (A.5) with (7) and (8) at $H_1 = \frac{1}{2}$.

*Timashev S.F., et al., Analysis of Biomedical Signals by Flicker-Noise Spectroscopy..., Las. Phys., 2009, 19 (4)*

FIGURE CAPTIONS

Fig. 1. Schematic of "random walk" evolution.

Fig. 2. Normalized functions $\varphi_1(x)$ (curve 1) and $\varphi_2(x)$ (curve 2).

Fig. 3. Sample setup for recording MEG signals and the scheme for placing SQUID-sensors.

Fig. 4. Analysis of the MEG signal recorded at sensor 10 for control subject 7 as the response to RB-stimulus ($T = 3.2$ s; $\sigma = 10.1$ fTl/cm, $H_1 = 1.27$, $T_1 = 2.9 \cdot 10^{-2}$ s, $D \approx 3.5 \cdot 10^3$ fTl$^2$/(cm$^2$ s), $S_{cS}(0) = 1.07 \cdot 10^4$ fTl$^2$ / (cm$^2$ $f_d$), $T_0 = 3.8 \cdot 10^{-2}$ s, $n_0 = 3.2$): (a) source signal; (b) power spectrum $S(f)$ given by Eq. (3) in the low-frequency range (main peaks: 1.6 – 9.4 – 22.8 – 27.8 Hz); (b) "experimental" – $\Phi^{(2)}(\tau)$ given by Eq. (4), "general interpolation" – $\Phi^{(2)}(\tau)$ given by Eq. (18), "resonant interpolation" – $\tilde{\Phi}_r^{(2)}(\tau)$ given by Eq. (17); (d) – "experimental – resonant" – $\Phi^{(2)}(\tau)$ given by Eq. (18) minus $\tilde{\Phi}_r^{(2)}(\tau)$ given by Eq. (17), "chaotic interpolation" – $\Phi_c^{(2)}(\tau)$ given by Eq. (6).

Fig. 5. Analysis of the MEG signal recorded at sensor 10 for control subject 9 as the response to RB-stimulus ($T = 3.2$ s; $\sigma = 10.1$ fTl/cm, $H_1 = 0.67$, $T_1 = 7.5 \cdot 10^{-2}$ s, $D \approx 1.36 \cdot 10^3$ fTl$^2$/(cm$^2$ s), $S_{cS}(0) = 1.74 \cdot 10^4$ fTl$^2$ / (cm$^2$ $f_d$), $T_0 = 7.4 \cdot 10^{-2}$ s, $n_0 = 2.2$): (a) source signal; (b) power spectrum $S(f)$ given by Eq. (3) in the low-frequency range (main peaks: 1.7 – 6 – 12.5 – 24.5 – 28.5 Hz); (c) "experimental" – $\Phi^{(2)}(\tau)$ given by Eq. (4), "general interpolation" – $\Phi^{(2)}(\tau)$ given by Eq. (18), "resonant interpolation" – $\tilde{\Phi}_r^{(2)}(\tau)$ given by Eq. (17); (d) – "experimental – resonant" – $\Phi^{(2)}(\tau)$ given by Eq. (18) minus $\tilde{\Phi}_r^{(2)}(\tau)$ given by Eq. (17), "chaotic interpolation" – $\Phi_c^{(2)}(\tau)$ given by Eq. (6).

Fig. 6. Analysis of the MEG signal recorded at sensor 10 for the PSE patient as the response to RB-stimulus ($T = 1.7$ s; $\sigma = 8.2$ fTl/cm, $H_1 = 7.1$, $T_1 = 2 \cdot 10^{-4}$ s, $D \sim 10^6$ fTl$^2$/(cm$^2$ s), $S_{cS}(0) = 3.79 \cdot 10^2$ fTl$^2$ / (cm$^2$ $f_d$), $T_0 = 2.1 \cdot 10^{-3}$ s, $n_0 = 3.04$): (a) source signal; (b) power spectrum $S(f)$ given by Eq. (3) in the low-frequency range (main peaks: 2.5 – 10 – 20 – 40 – 50 – 60 – 100 Hz); (c) "experimental" – $\Phi^{(2)}(\tau)$ given by Eq. (4), "general interpolation" – $\Phi^{(2)}(\tau)$ given by Eq. (18), "resonant interpolation" – $\tilde{\Phi}_r^{(2)}(\tau)$ given by Eq. (17); (d) – "experimental – resonant" –

$\Phi^{(2)}(\tau)$ given by Eq. (18) minus $\tilde{\Phi}_r^{(2)}(\tau)$ given by Eq. (17), "chaotic interpolation" – $\Phi_c^{(2)}(\tau)$ given by Eq. (6).

Fig. 7. Analysis of the MEG signal recorded at sensor 10 for control subject 7 as the response to RG-stimulus ($T = 3.2$ s; $\sigma = 10.4$ fTl/cm, $H_1 = 0.81$, $T_1 = 0.12$ s, $D \approx 9.01 \cdot 10^2$ fTl$^2$/(cm$^2$ s), $S_{cS}(0) = 4.3 \cdot 10^3$ fTl$^2$ / (cm$^2$ $f_d$), $T_0 = 0.147$ s, $n_0 = 2.3$): (a) source signal; (b) power spectrum $S(f)$ given by Eq. (3) in the low-frequency range (main peaks: 1.9 – 3.5 – 4.5 – 9.5 – 27.8 Hz); (c) "experimental" – $\Phi^{(2)}(\tau)$ given by Eq. (4), "general interpolation" – $\Phi^{(2)}(\tau)$ given by Eq. (18), "resonant interpolation" – $\tilde{\Phi}_r^{(2)}(\tau)$ given by Eq. (17); (d) – "experimental – resonant" – $\Phi^{(2)}(\tau)$ given by Eq. (18) minus $\tilde{\Phi}_r^{(2)}(\tau)$ given by Eq. (17), "chaotic interpolation" – $\Phi_c^{(2)}(\tau)$ given by Eq. (6).

Fig. 8. Analysis of the MEG signal recorded at sensor 10 for control subject 9 as the response to RG-stimulus ($T = 3.2$ s; $\sigma = 15.3$ fTl/cm, $H_1 = 0.5$, $T_1 = 0.24$ s, $D \approx 9.75 \cdot 10^2$ fTl$^2$/(cm$^2$ s), $S_{cS}(0) = 6.6 \cdot 10^4$ fTl$^2$ / (cm$^2$ $f_d$), $T_0 = 0.145$ s, $n_0 = 2.14$): (a) source signal; (b) power spectrum $S(f)$ given by Eq. (3) in the low-frequency range (main peaks: 3.8 – 5.7 – 7.6 – 10.7 – 18.3 Hz); (c) "experimental" – $\Phi^{(2)}(\tau)$ given by Eq. (4), "general interpolation" – $\Phi^{(2)}(\tau)$ given by Eq. (18), "resonant interpolation" – $\tilde{\Phi}_r^{(2)}(\tau)$ given by Eq. (17); (d) – "experimental – resonant" – $\Phi^{(2)}(\tau)$ given by Eq. (18) minus $\tilde{\Phi}_r^{(2)}(\tau)$ given by Eq. (17), "chaotic interpolation" – $\Phi_c^{(2)}(\tau)$ given by Eq. (6).

Fig. 9. Analysis of the MEG signal recorded at sensor 10 for the PSE patient as the response to RG-stimulus ($T = 1.7$ s; $\sigma = 10.6$ fTl/cm, $H_1 = 0.12$, $T_1 = 4.97 \cdot 10^{-2}$ s, $D \sim 2.25 \cdot 10^3$ fTl$^2$/(cm$^2$ s), $S_{cS}(0) = 0.55 \cdot 10^4$ fTl$^2$ / (cm$^2$ $f_d$), $T_0 = 5.9 \cdot 10^{-2}$ s, $n_0 = 1.07$): (a) source signal; (b) power spectrum $S(f)$ given by Eq. (3) in the low-frequency range (main peaks: 4.1 – 40.6 – 50 – 59.5 – 100 Hz); (c) "experimental" – $\Phi^{(2)}(\tau)$ given by Eq. (4), "general interpolation" – $\Phi^{(2)}(\tau)$ given by Eq. (18), "resonant interpolation" – $\tilde{\Phi}_r^{(2)}(\tau)$ given by Eq. (17); (d) – "experimental – resonant" – $\Phi^{(2)}(\tau)$ given by Eq. (18) minus $\tilde{\Phi}_r^{(2)}(\tau)$ given by Eq. (17), "chaotic interpolation" – $\Phi_c^{(2)}(\tau)$ given by Eq. (6).

Fig. 10. Analysis of the MEG signal recorded at sensor 43 for control subject 7 as the response to RB-stimulus ($T = 3.2$ s; $\sigma = 22.2$ fTl/cm, $H_1 = 1.24$, $T_1 = 3.52 \cdot 10^{-2}$ s, $D \approx 1.4 \cdot 10^4$ fTl$^2$/(cm$^2$ s), $S_{cS}(0) = 6.1 \cdot 10^4$ fTl$^2$ / (cm$^2$ $f_d$), $T_0 = 4.4 \cdot 10^{-2}$ s, $n_0 = 3.3$): (a) source signal; (b) power

spectrum $S(f)$ given by Eq. (3) in the low-frequency range (main peaks: 1.6 – 12.6 – 19.1 – 37.5 Hz); (c) "experimental" – $\Phi^{(2)}(\tau)$ given by Eq. (4), "general interpolation" – $\Phi^{(2)}(\tau)$ given by Eq. (18), "resonant interpolation" – $\tilde{\Phi}_r^{(2)}(\tau)$ given by Eq. (17); (d) – "experimental – resonant" – $\Phi^{(2)}(\tau)$ given by Eq. (18) minus $\tilde{\Phi}_r^{(2)}(\tau)$ given by Eq. (17), "chaotic interpolation" – $\Phi_c^{(2)}(\tau)$ given by Eq. (6).

Fig. 11. Analysis of the MEG signal recorded at sensor 43 for control subject 9 as the response to RB-stimulus ($T = 3.2$ s; $\sigma = 9.7$ fTl/cm, $H_1 = 0.51$, $T_1 = 8\cdot10^{-2}$ s, $D \approx 1.18\cdot10^3$ fTl$^2$/(cm$^2$ s), $S_{cS}(0) = 1.03\cdot10^4$ fTl$^2$ / (cm$^2$ $f_d$), $T_0 = 5.28\cdot10^{-2}$ s, $n_0 = 2.2$): (a) source signal; (b) power spectrum $S(f)$ given by Eq. (3) in the low-frequency range (main peaks: 2.5 – 10 – 17.5 – 28 – 37.5 Hz); (c) "experimental" – $\Phi^{(2)}(\tau)$ given by Eq. (4), "general interpolation" – $\Phi^{(2)}(\tau)$ given by Eq. (18), "resonant interpolation" – $\tilde{\Phi}_r^{(2)}(\tau)$ given by Eq. (17); (d) – "experimental – resonant" – $\Phi^{(2)}(\tau)$ given by Eq. (18) minus $\tilde{\Phi}_r^{(2)}(\tau)$ given by Eq. (17), "chaotic interpolation" – $\Phi_c^{(2)}(\tau)$ given by Eq. (6).

Fig. 12. Analysis of the MEG signal recorded at sensor 43 for the PSE patient as the response to RB-stimulus ($T = 1.7$ s; $\sigma = 11.1$ fTl/cm, $H_1 = 0.31$, $T_1 = 3.6\cdot10^{-2}$ s, $D \approx 3.4\cdot10^3$ fTl$^2$/(cm$^2$ s), $S_{cS}(0) = 1.6\cdot10^4$ fTl$^2$ / (cm$^2$ $f_d$), $T_0 = 7.88\cdot10^{-2}$ s, $n_0 = 1.46$): (a) source signal; (b) power spectrum $S(f)$ given by Eq. (3) in the low-frequency range (main peaks: 1.6 – 11.5 – 50 Hz); (c) "experimental" – $\Phi^{(2)}(\tau)$ given by Eq. (4), "general interpolation" – $\Phi^{(2)}(\tau)$ given by Eq. (18), "resonant interpolation" – $\tilde{\Phi}_r^{(2)}(\tau)$ given by Eq. (17); (d) – "experimental – resonant" – $\Phi^{(2)}(\tau)$ given by Eq. (18) minus $\tilde{\Phi}_r^{(2)}(\tau)$ given by Eq. (17), "chaotic interpolation" – $\Phi_c^{(2)}(\tau)$ given by Eq. (6).

Fig. 13. Analysis of the MEG signal recorded at sensor 43 for control subject 7 as the response to RG-stimulus ($T = 3.2$ s; $\sigma = 16$ fTl/cm, $H_1 = 0.94$, $T_1 = 7.74\cdot10^{-2}$ s, $D \approx 3.31\cdot10^3$ fTl$^2$/(cm$^2$ s), $S_{cS}(0) = 5.03\cdot10^4$ fTl$^2$ / (cm$^2$ $f_d$), $T_0 = 8.28\cdot10^{-2}$ s, $n_0 = 2.6$): (a) source signal; (b) power spectrum $S(f)$ given by Eq. (3) in the low-frequency range (main peaks: 3.1 – 9.1 – 10.9 – 16.2 – 18.7 – 28.4 Hz); (c) "experimental" – $\Phi^{(2)}(\tau)$ given by Eq. (4), "general interpolation" – $\Phi^{(2)}(\tau)$ given by Eq. (18), "resonant interpolation" – $\tilde{\Phi}_r^{(2)}(\tau)$ given by Eq. (17); (d) – "experimental – resonant" – $\Phi^{(2)}(\tau)$ given by Eq. (18) minus $\tilde{\Phi}_r^{(2)}(\tau)$ given by Eq. (17), "chaotic interpolation" – $\Phi_c^{(2)}(\tau)$ given by Eq. (6).

Fig. 14. Analysis of the MEG signal recorded at sensor 43 for control subject 9 as the response to RG-stimulus ($T$ = 3.2 s; $\sigma$ = 14.4 fTl/cm, $H_1$ = 1.44, $T_1$ = 2.04·10$^{-2}$ s, $D \approx$ 1.02·10$^4$ fTl$^2$/(cm$^2$ s), $S_{cS}(0)$ = 1.34·10$^4$ fTl$^2$ / (cm$^2$ $f_d$), $T_0$ = 2.42·10$^{-2}$ s, $n_0$ = 4): (a) source signal; (b) power spectrum $S(f)$ given by Eq. (3) in the low-frequency range (main peaks: 10 – 19 Hz); (c) "experimental" – $\Phi^{(2)}(\tau)$ given by Eq. (4), "general interpolation" – $\Phi^{(2)}(\tau)$ given by Eq. (18), "resonant interpolation" – $\tilde{\Phi}_r^{(2)}(\tau)$ given by Eq. (17); (d) – "experimental – resonant" – $\Phi^{(2)}(\tau)$ given by Eq. (18) minus $\tilde{\Phi}_r^{(2)}(\tau)$ given by Eq. (17), "chaotic interpolation" – $\Phi_c^{(2)}(\tau)$ given by Eq. (6).

Fig. 15. Analysis of the MEG signal recorded at sensor 43 for the PSE patient as the response to RG-stimulus ($T$ = 1.7 s; $\sigma$ = 20.2 fTl/cm, $H_1$ = 0.47, $T_1$ = 3.2·10$^{-2}$ s, $D \approx$ 1.27·10$^4$ fTl$^2$/(cm$^2$ s), $S_{cS}(0)$ = 1.73·10$^4$ fTl$^2$ / (cm$^2$ $f_d$), $T_0$ = 1.97·10$^{-2}$ s, $n_0$ = 2.2): (a) source signal; (b) power spectrum $S(f)$ given by Eq. (3) in the low-frequency range (main peaks: 2.5 – 6 – 17.9 – 50 Hz); (c) "experimental" – $\Phi^{(2)}(\tau)$ given by Eq. (4), "general interpolation" – $\Phi^{(2)}(\tau)$ given by Eq. (18), "resonant interpolation" – $\tilde{\Phi}_r^{(2)}(\tau)$ given by Eq. (17); (d) – "experimental – resonant" – $\Phi^{(2)}(\tau)$ given by Eq. (18) minus $\tilde{\Phi}_r^{(2)}(\tau)$ given by Eq. (17), "chaotic interpolation" – $\Phi_c^{(2)}(\tau)$ given by Eq. (6).

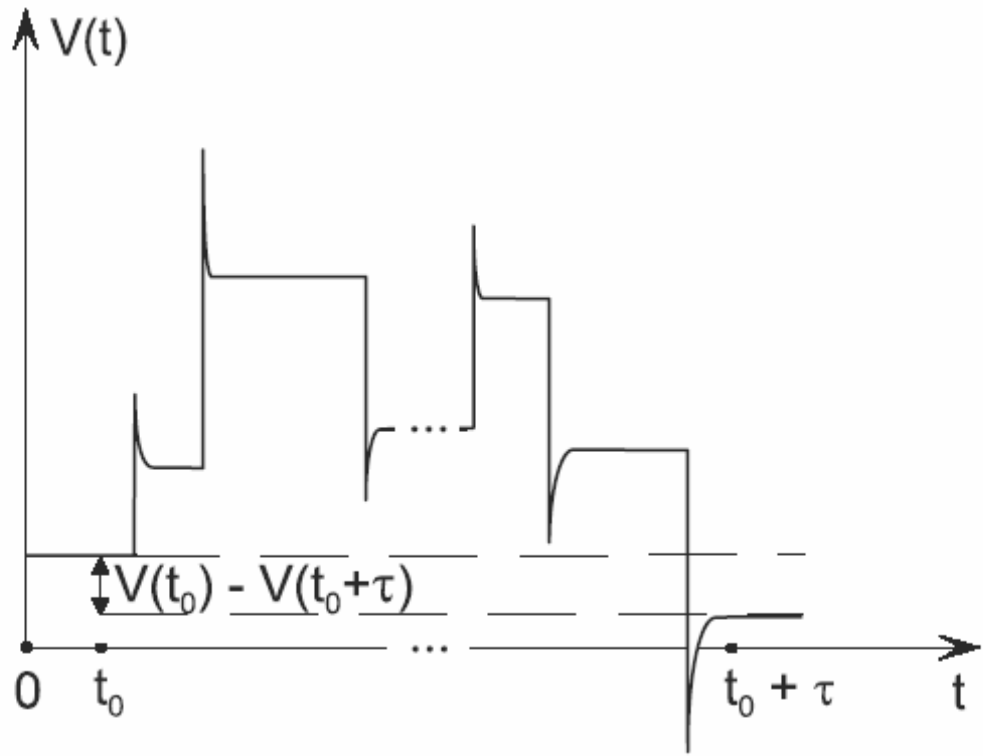

Fig. 1. Schematic of "random walk" evolution.

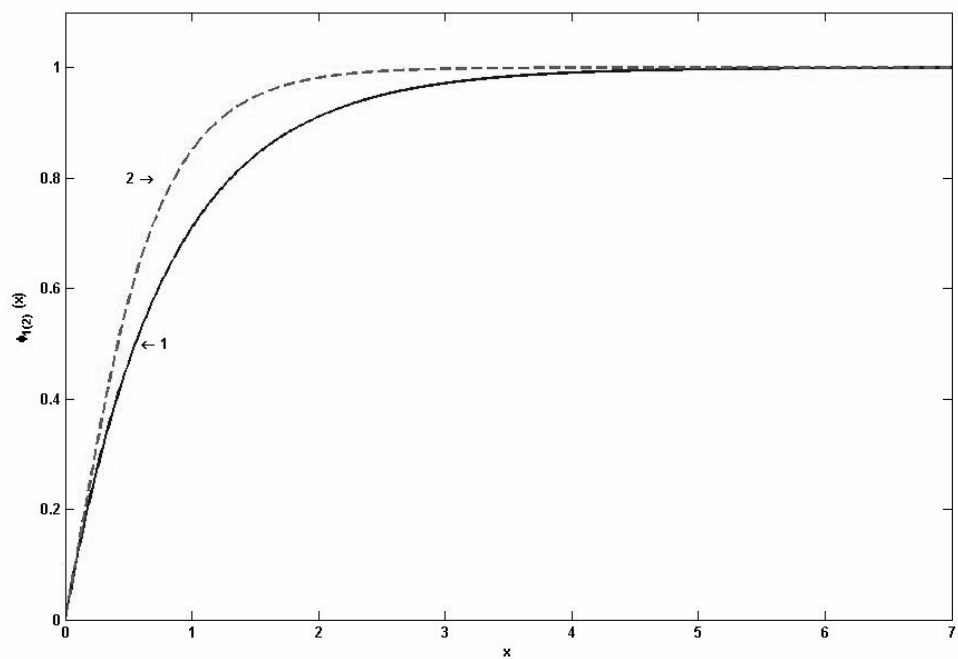

Fig. 2. Normalized functions $\varphi_1(x)$ (curve 1) and $\varphi_2(x)$ (curve 2).

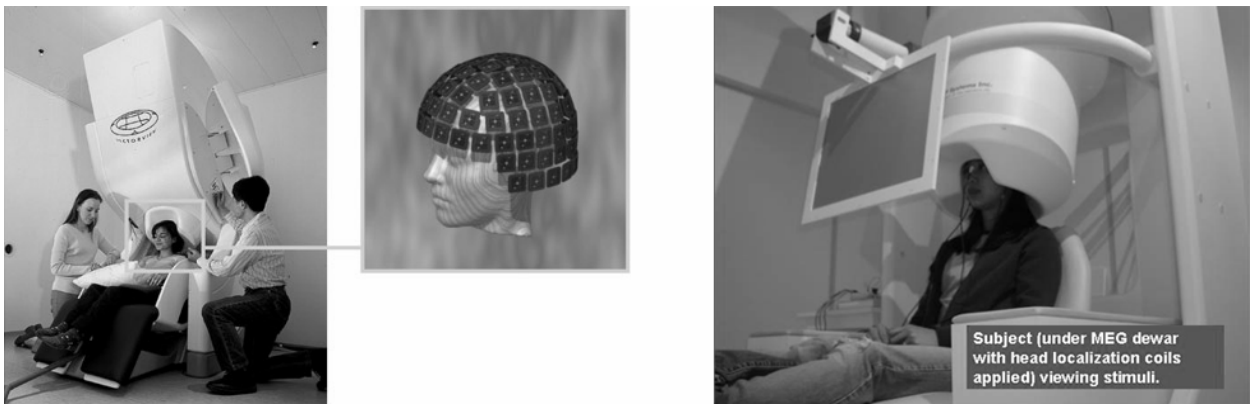

Fig. 3. Sample setup for recording MEG signals and the scheme for placing SQUID-sensors.

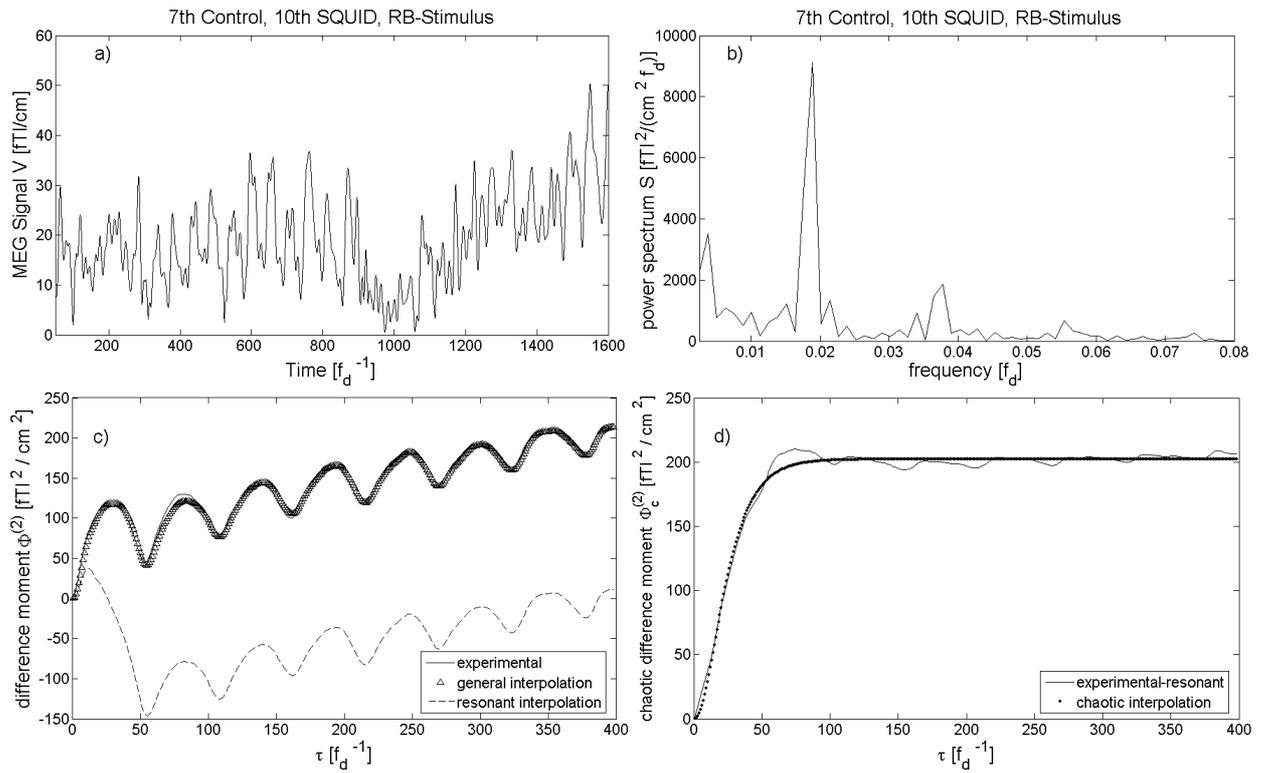

Fig. 4. Analysis of the MEG signal recorded at sensor 10 for control subject 7 as the response to RB-stimulus ($T = 3.2$ s; $\sigma = 10.1$ fTl/cm, $H_1 = 1.27$, $T_1 = 2.9 \cdot 10^{-2}$ s, $D \approx 3.5 \cdot 10^3$ fTl$^2$/(cm$^2$ s), $S_{cS}(0) = 1.07 \cdot 10^4$ fTl$^2$ / (cm$^2$ $f_d$), $T_0 = 3.8 \cdot 10^{-2}$ s, $n_0 = 3.2$): (a) source signal; (b) power spectrum $S(f)$ given by Eq. (3) in the low-frequency range (main peaks: 1.6 – 9.4 –22.8 – 27.8 Hz); (b) "experimental" – $\Phi^{(2)}(\tau)$ given by Eq. (4), "general interpolation" – $\Phi^{(2)}(\tau)$ given by Eq. (18), "resonant interpolation" – $\tilde{\Phi}_r^{(2)}(\tau)$ given by Eq. (17); (d) – "experimental – resonant" – $\Phi^{(2)}(\tau)$ given by Eq. (18) minus $\tilde{\Phi}_r^{(2)}(\tau)$ given by Eq. (17), "chaotic interpolation" – $\Phi_c^{(2)}(\tau)$ given by Eq. (6).

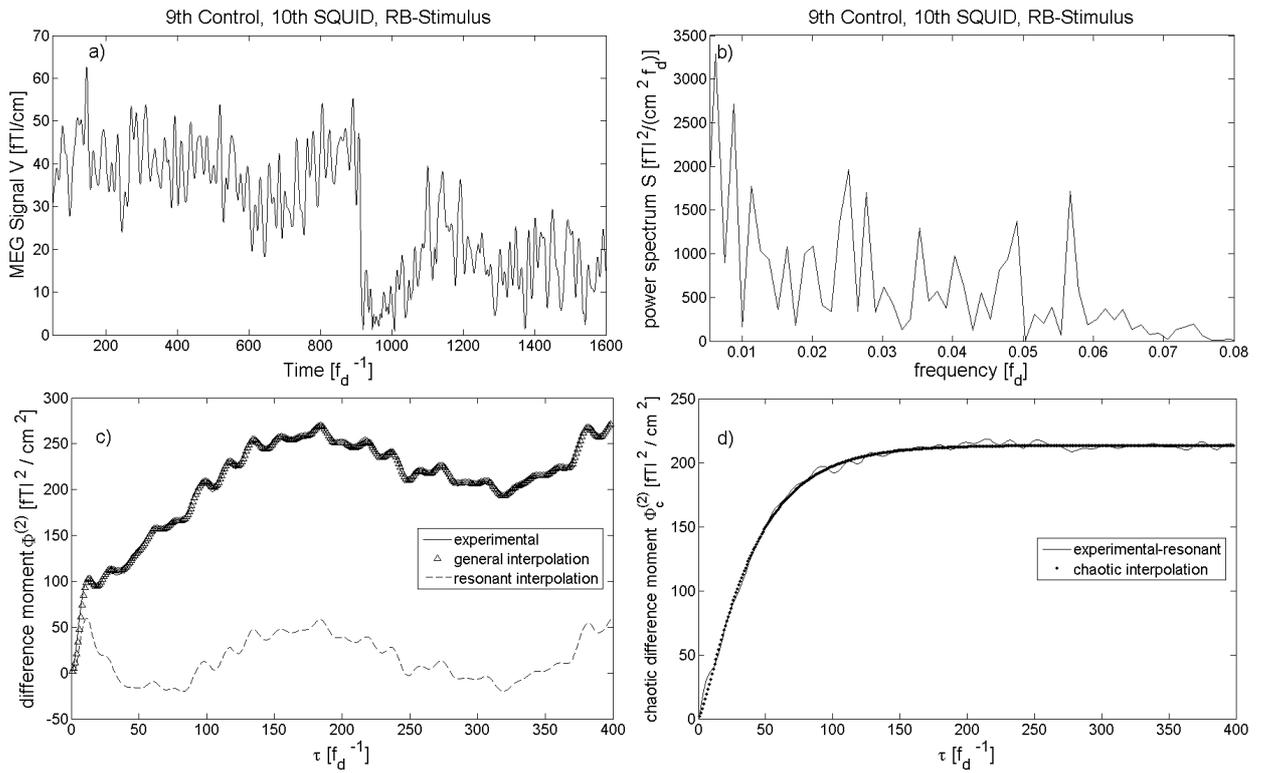

Fig. 5. Analysis of the MEG signal recorded at sensor 10 for control subject 9 as the response to RB-stimulus ($T$ = 3.2 s; $\sigma$ =10.1 fTl/cm, $H_1$ = 0.67, $T_1$ = 7.5·10$^{-2}$ s, $D \approx$ 1.36·10$^3$ fTl$^2$/(cm$^2$ s), $S_{cS}(0)$ = 1.74·10$^4$ fTl$^2$ / (cm$^2$ $f_d$), $T_0$ = 7.4·10$^{-2}$ s, $n_0$ = 2.2): (a) source signal; (b) power spectrum $S(f)$ given by Eq. (3) in the low-frequency range (main peaks: 1.7 – 6 – 12.5 – 24.5 – 28.5 Hz); (c) "experimental" – $\Phi^{(2)}(\tau)$ given by Eq. (4), "general interpolation" – $\Phi^{(2)}(\tau)$ given by Eq. (18), "resonant interpolation" – $\tilde{\Phi}_r^{(2)}(\tau)$ given by Eq. (17); (d) – "experimental – resonant" – $\Phi^{(2)}(\tau)$ given by Eq. (18) minus $\tilde{\Phi}_r^{(2)}(\tau)$ given by Eq. (17), "chaotic interpolation" – $\Phi_c^{(2)}(\tau)$ given by Eq. (6).

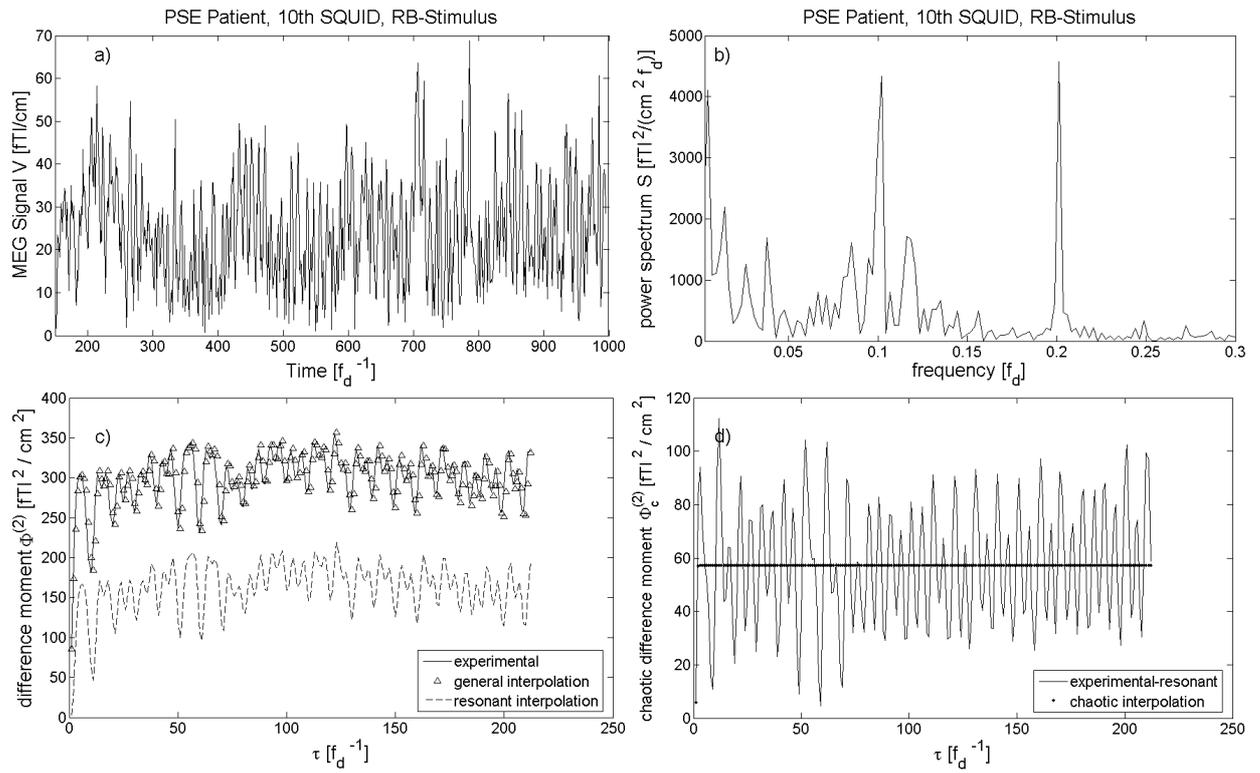

Fig. 6. Analysis of the MEG signal recorded at sensor 10 for the PSE patient as the response to RB-stimulus ($T = 1.7$ s; $\sigma = 8.2$ fTl/cm, $H_1 = 7.1$, $T_1 = 2\cdot10^{-4}$ s, $D \sim 10^6$ fTl$^2$/(cm$^2$ s), $S_{cS}(0) = 3.79\cdot10^2$ fTl$^2$/(cm$^2 f_d$), $T_0 = 2.1\cdot10^{-3}$ s, $n_0 = 3.04$): (a) source signal; (b) power spectrum $S(f)$ given by Eq. (3) in the low-frequency range (main peaks: 2.5 – 10 – 20 – 40 – 50 – 60 – 100 Hz); (c) "experimental" – $\Phi^{(2)}(\tau)$ given by Eq. (4), "general interpolation" – $\Phi^{(2)}(\tau)$ given by Eq. (18), "resonant interpolation" – $\tilde{\Phi}_r^{(2)}(\tau)$ given by Eq. (17); (d) – "experimental – resonant" – $\Phi^{(2)}(\tau)$ given by Eq. (18) minus $\tilde{\Phi}_r^{(2)}(\tau)$ given by Eq. (17), "chaotic interpolation" – $\Phi_c^{(2)}(\tau)$ given by Eq. (6).

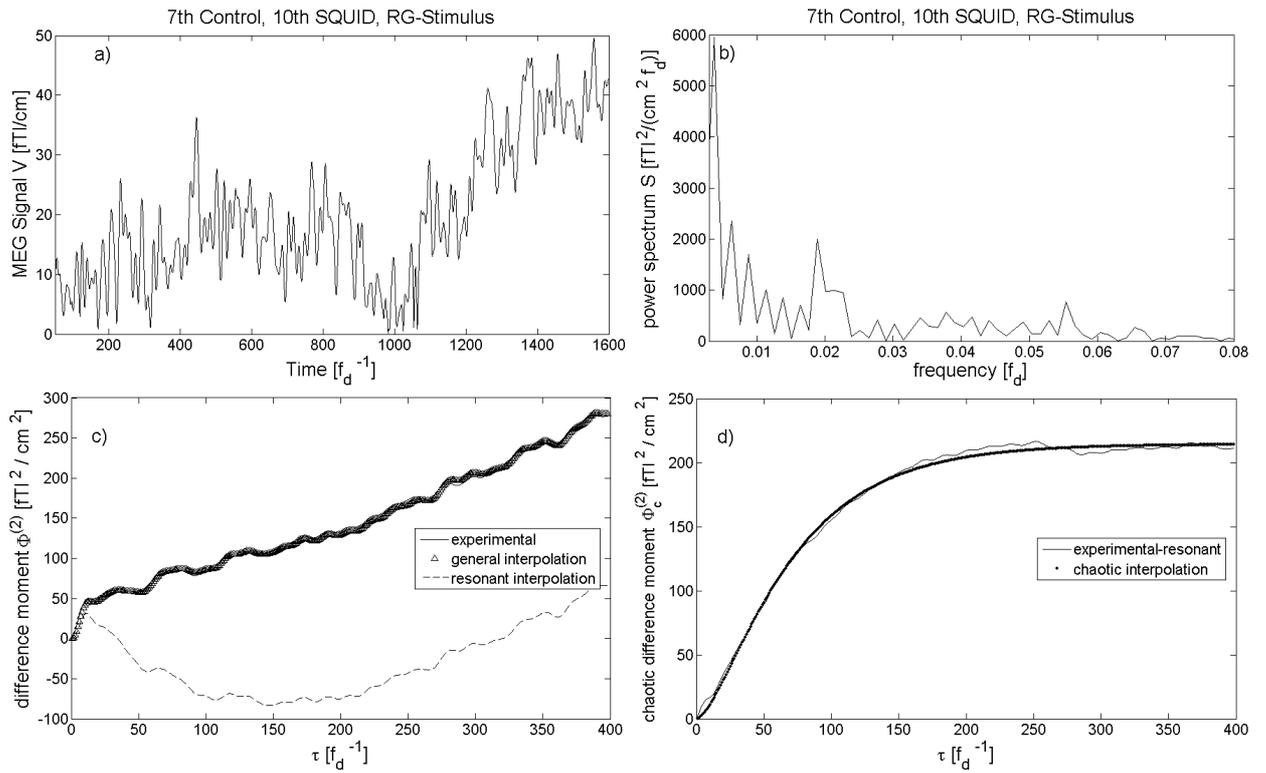

Fig. 7. Analysis of the MEG signal recorded at sensor 10 for control subject 7 as the response to RG-stimulus ($T = 3.2$ s; $\sigma = 10.4$ fTl/cm, $H_1 = 0.81$, $T_1 = 0.12$ s, $D \approx 9.01 \cdot 10^{2}$ fTl$^2$/(cm$^2$ s), $S_{cS}(0) = 4.3 \cdot 10^3$ fTl$^2$ / (cm$^2$ $f_d$), $T_0 = 0.147$ s, $n_0 = 2.3$): (a) source signal; (b) power spectrum $S(f)$ given by Eq. (3) in the low-frequency range (main peaks: 1.9 – 3.5 – 4.5 – 9.5 – 27.8 Hz); (c) "experimental" – $\Phi^{(2)}(\tau)$ given by Eq. (4), "general interpolation" – $\Phi^{(2)}(\tau)$ given by Eq. (18), "resonant interpolation" – $\tilde{\Phi}_r^{(2)}(\tau)$ given by Eq. (17); (d) – "experimental – resonant" – $\Phi^{(2)}(\tau)$ given by Eq. (18) minus $\tilde{\Phi}_r^{(2)}(\tau)$ given by Eq. (17), "chaotic interpolation" – $\Phi_c^{(2)}(\tau)$ given by Eq. (6).

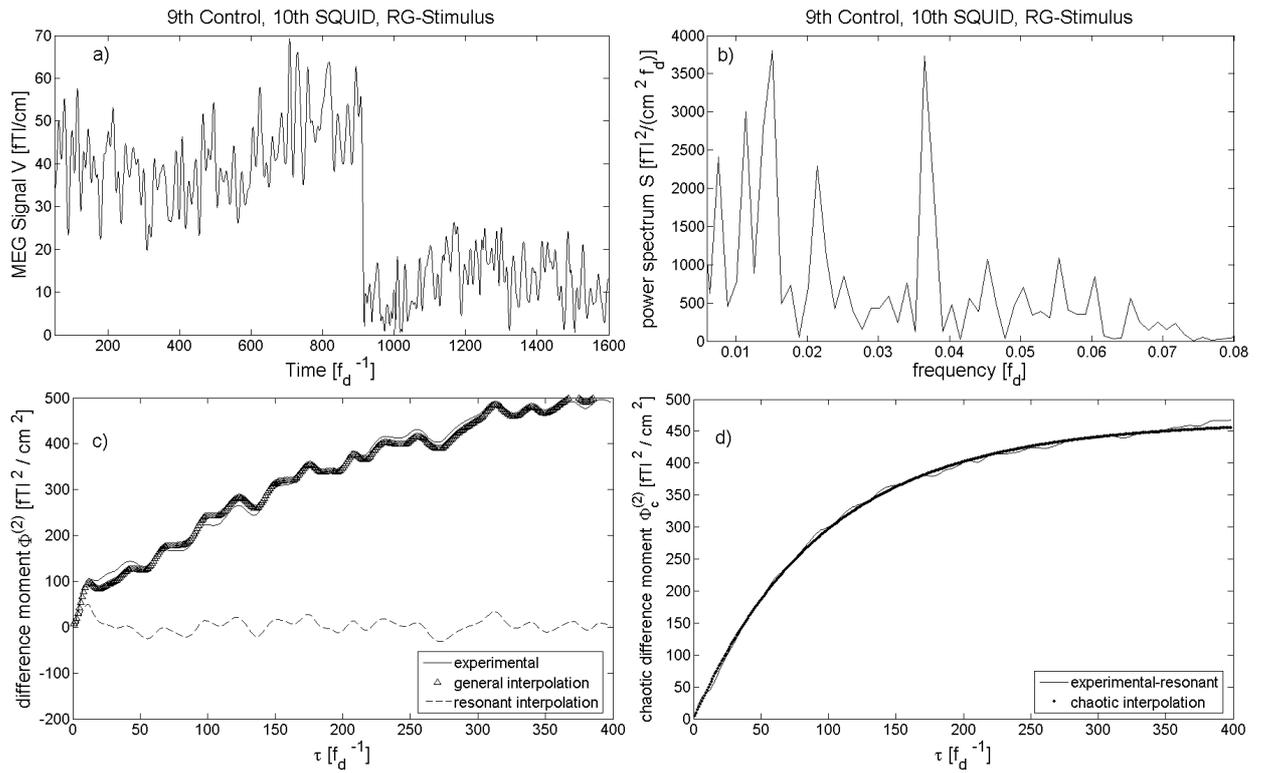

Fig. 8. Analysis of the MEG signal recorded at sensor 10 for control subject 9 as the response to RG-stimulus ($T$ = 3.2 s; $\sigma$ = 15.3 fTl/cm, $H_1$ = 0.5, $T_1$ = 0.24 s, $D \approx 9.75 \cdot 10^2$ fTl$^2$/(cm$^2$ s), $S_{cS}(0)$ = $6.6 \cdot 10^4$ fTl$^2$ / (cm$^2$ $f_d$), $T_0$ = 0.145 s, $n_0$ = 2.14): (a) source signal; (b) power spectrum $S(f)$ given by Eq. (3) in the low-frequency range (main peaks: 3.8 – 5.7 – 7.6 – 10.7 – 18.3 Hz); (c) "experimental" – $\Phi^{(2)}(\tau)$ given by Eq. (4), "general interpolation" – $\Phi^{(2)}(\tau)$ given by Eq. (18), "resonant interpolation" – $\tilde{\Phi}_r^{(2)}(\tau)$ given by Eq. (17); (d) – "experimental – resonant" – $\Phi^{(2)}(\tau)$ given by Eq. (18) minus $\tilde{\Phi}_r^{(2)}(\tau)$ given by Eq. (17), "chaotic interpolation" – $\Phi_c^{(2)}(\tau)$ given by Eq. (6).

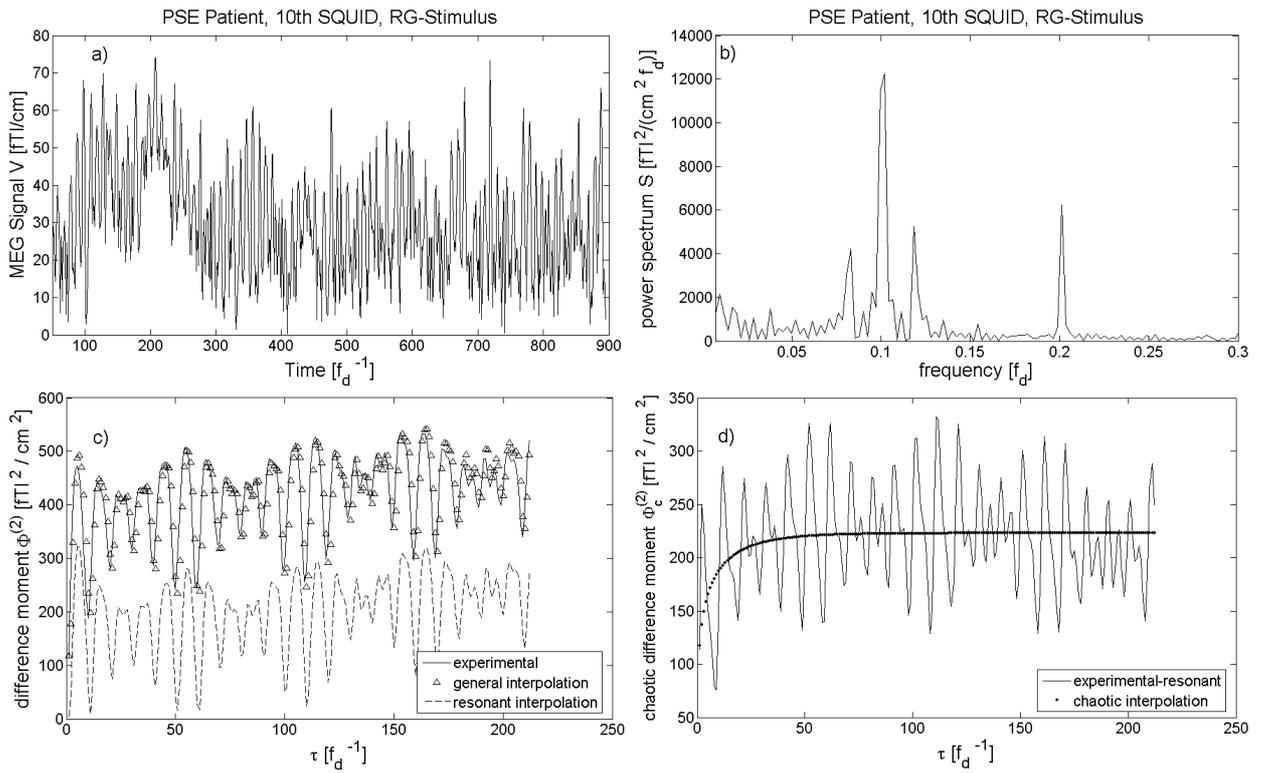

Fig. 9. Analysis of the MEG signal recorded at sensor 10 for the PSE patient as the response to RG-stimulus ($T = 1.7$ s; $\sigma = 10.6$ fTl/cm, $H_1 = 0.12$, $T_1 = 4.97 \cdot 10^{-2}$ s, $D \sim 2.25 \cdot 10^3$ fTl$^2$/(cm$^2$ s), $S_{cS}(0) = 0.55 \cdot 10^4$ fTl$^2$ / (cm$^2$ $f_d$), $T_0 = 5.9 \cdot 10^{-2}$ s, $n_0 = 1.07$): (a) source signal; (b) power spectrum $S(f)$ given by Eq. (3) in the low-frequency range (main peaks: 4.1 – 40.6 – 50 – 59.5 – 100 Hz); (c) "experimental" – $\Phi^{(2)}(\tau)$ given by Eq. (4), "general interpolation" – $\Phi^{(2)}(\tau)$ given by Eq. (18), "resonant interpolation" – $\tilde{\Phi}_r^{(2)}(\tau)$ given by Eq. (17); (d) – "experimental – resonant" – $\Phi^{(2)}(\tau)$ given by Eq. (18) minus $\tilde{\Phi}_r^{(2)}(\tau)$ given by Eq. (17), "chaotic interpolation" – $\Phi_c^{(2)}(\tau)$ given by Eq. (6).

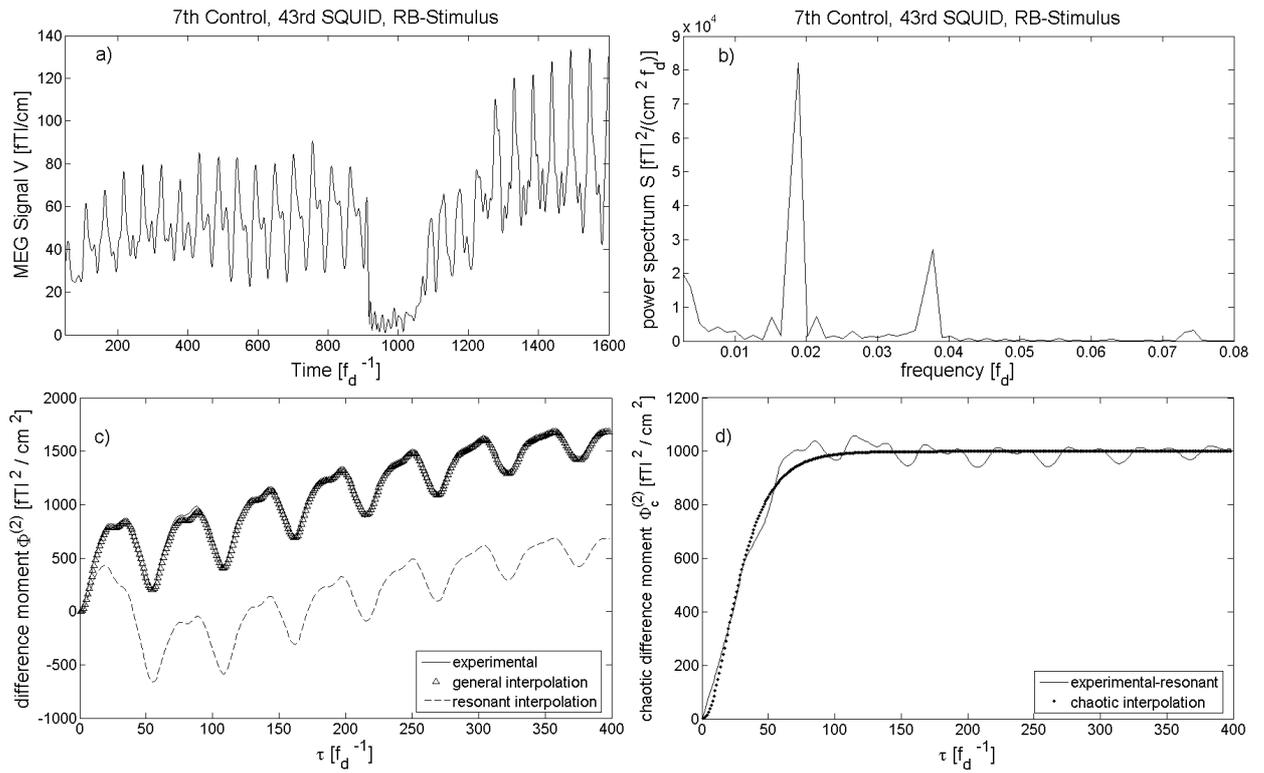

Fig. 10. Analysis of the MEG signal recorded at sensor 43 for control subject 7 as the response to RB-stimulus ($T = 3.2$ s; $\sigma = 22.2$ fTl/cm, $H_1 = 1.24$, $T_1 = 3.52 \cdot 10^{-2}$ s, $D \approx 1.4 \cdot 10^4$ fTl$^2$/(cm$^2$ s), $S_{cS}(0) = 6.1 \cdot 10^4$ fTl$^2$/(cm$^2$ $f_d$), $T_0 = 4.4 \cdot 10^{-2}$ s, $n_0 = 3.3$): (a) source signal; (b) power spectrum $S(f)$ given by Eq. (3) in the low-frequency range (main peaks: 1.6 – 12.6 – 19.1 – 37.5 Hz); (c) "experimental" – $\Phi^{(2)}(\tau)$ given by Eq. (4), "general interpolation" – $\Phi^{(2)}(\tau)$ given by Eq. (18), "resonant interpolation" – $\tilde{\Phi}_r^{(2)}(\tau)$ given by Eq. (17); (d) – "experimental – resonant" – $\Phi^{(2)}(\tau)$ given by Eq. (18) minus $\tilde{\Phi}_r^{(2)}(\tau)$ given by Eq. (17), "chaotic interpolation" – $\Phi_c^{(2)}(\tau)$ given by Eq. (6).

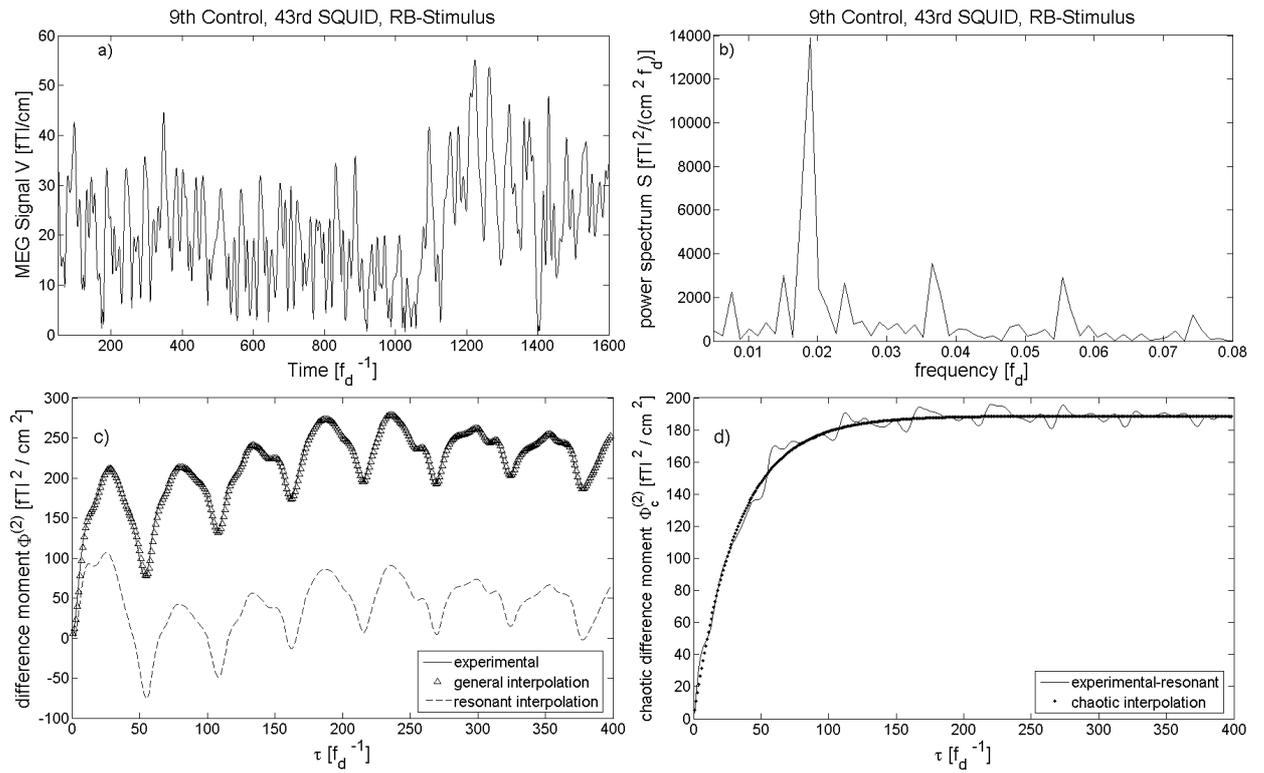

Fig. 11. Analysis of the MEG signal recorded at sensor 43 for control subject 9 as the response to RB-stimulus ($T = 3.2$ s; $\sigma = 9.7$ fTl/cm, $H_1 = 0.51$, $T_1 = 8 \cdot 10^{-2}$ s, $D \approx 1.18 \cdot 10^3$ fTl$^2$/(cm$^2$ s), $S_{cS}(0) = 1.03 \cdot 10^4$ fTl$^2$ / (cm$^2$ $f_d$), $T_0 = 5.28 \cdot 10^{-2}$ s, $n_0 = 2.2$): (a) source signal; (b) power spectrum $S(f)$ given by Eq. (3) in the low-frequency range (main peaks: 2.5 – 10 – 17.5 – 28 – 37.5 Hz); (c) "experimental" – $\Phi^{(2)}(\tau)$ given by Eq. (4), "general interpolation" – $\Phi^{(2)}(\tau)$ given by Eq. (18), "resonant interpolation" – $\tilde{\Phi}_r^{(2)}(\tau)$ given by Eq. (17); (d) – "experimental – resonant" – $\Phi^{(2)}(\tau)$ given by Eq. (18) minus $\tilde{\Phi}_r^{(2)}(\tau)$ given by Eq. (17), "chaotic interpolation" – $\Phi_c^{(2)}(\tau)$ given by Eq. (6).

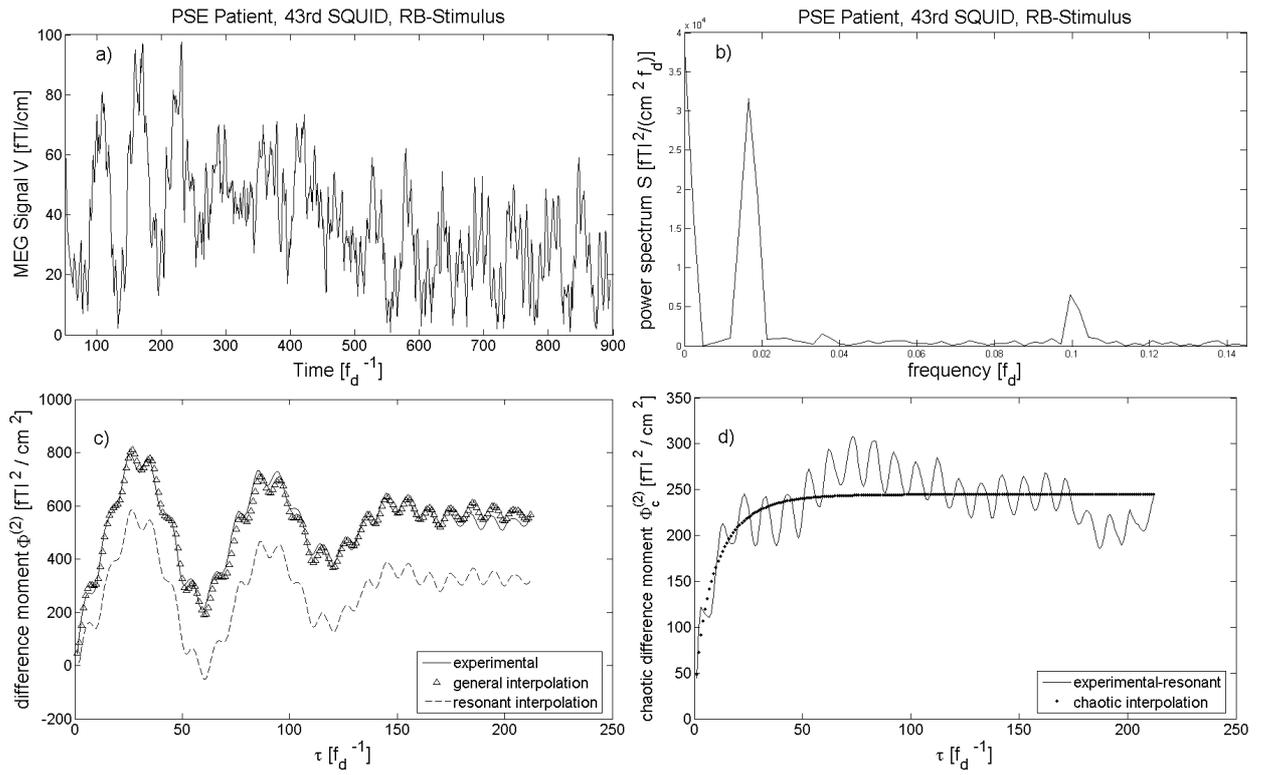

Fig. 12. Analysis of the MEG signal recorded at sensor 43 for the PSE patient as the response to RB-stimulus ($T = 1.7$ s; $\sigma = 11.1$ fTl/cm, $H_1 = 0.31$, $T_1 = 3.6 \cdot 10^{-2}$ s, $D \approx 3.4 \cdot 10^3$ fTl$^2$/(cm$^2$ s), $S_{cS}(0) = 1.6 \cdot 10^4$ fTl$^2$ / (cm$^2$ $f_d$), $T_0 = 7.88 \cdot 10^{-2}$ s, $n_0 = 1.46$): (a) source signal; (b) power spectrum $S(f)$ given by Eq. (3) in the low-frequency range (main peaks: 1.6 – 11.5 – 50 Hz); (c) "experimental" – $\Phi^{(2)}(\tau)$ given by Eq. (4), "general interpolation" – $\Phi^{(2)}(\tau)$ given by Eq. (18), "resonant interpolation" – $\tilde{\Phi}_r^{(2)}(\tau)$ given by Eq. (17); (d) – "experimental – resonant" – $\Phi^{(2)}(\tau)$ given by Eq. (18) minus $\tilde{\Phi}_r^{(2)}(\tau)$ given by Eq. (17), "chaotic interpolation" – $\Phi_c^{(2)}(\tau)$ given by Eq. (6).

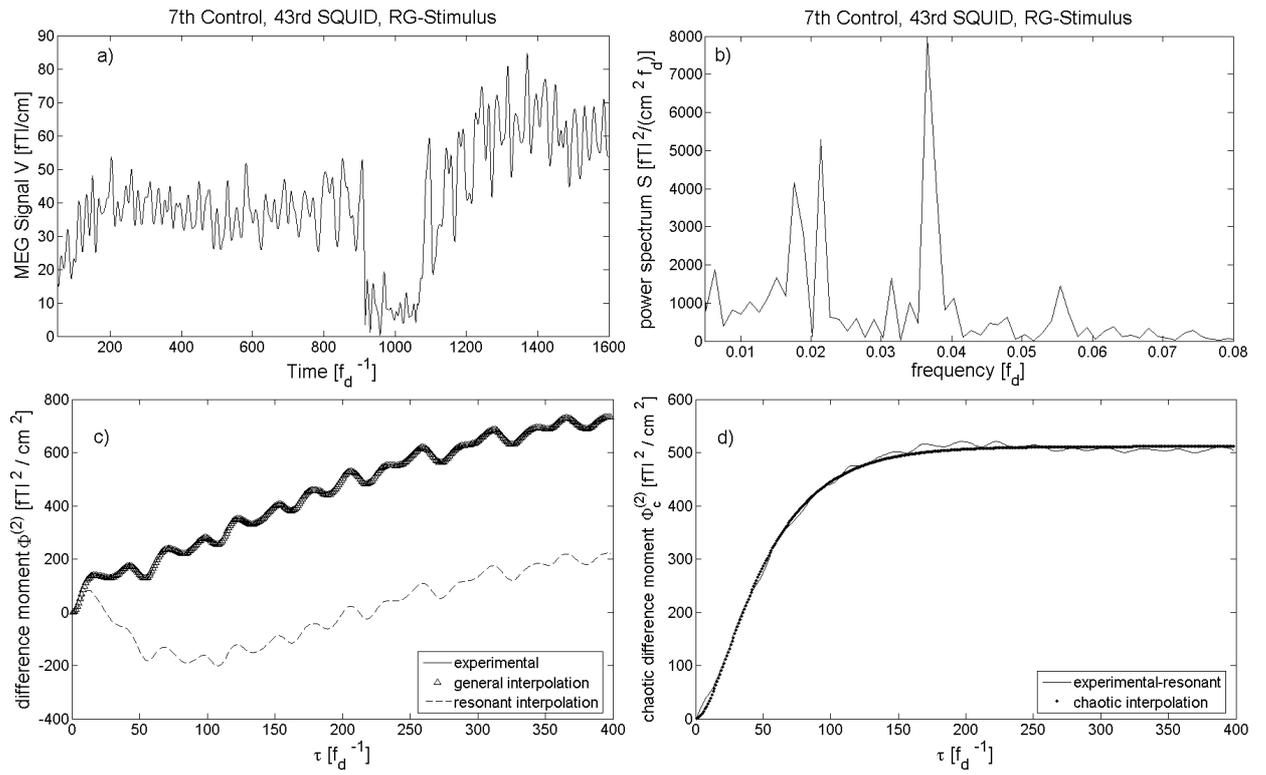

Fig. 13. Analysis of the MEG signal recorded at sensor 43 for control subject 7 as the response to RG-stimulus ($T = 3.2$ s; $\sigma = 16$ fTl/cm, $H_1 = 0.94$, $T_1 = 7.74 \cdot 10^{-2}$ s, $D \approx 3.31 \cdot 10^3$ fTl$^2$/(cm$^2$ s), $S_{cS}(0) = 5.03 \cdot 10^4$ fTl$^2$ / (cm$^2$ $f_d$), $T_0 = 8.28 \cdot 10^{-2}$ s, $n_0 = 2.6$): (a) source signal; (b) power spectrum $S(f)$ given by Eq. (3) in the low-frequency range (main peaks: 3.1 – 9.1 – 10.9 – 16.2 – 18.7 – 28.4 Hz); (c) "experimental" – $\Phi^{(2)}(\tau)$ given by Eq. (4), "general interpolation" – $\Phi^{(2)}(\tau)$ given by Eq. (18), "resonant interpolation" – $\tilde{\Phi}_r^{(2)}(\tau)$ given by Eq. (17); (d) – "experimental – resonant" – $\Phi^{(2)}(\tau)$ given by Eq. (18) minus $\tilde{\Phi}_r^{(2)}(\tau)$ given by Eq. (17), "chaotic interpolation" – $\Phi_c^{(2)}(\tau)$ given by Eq. (6).

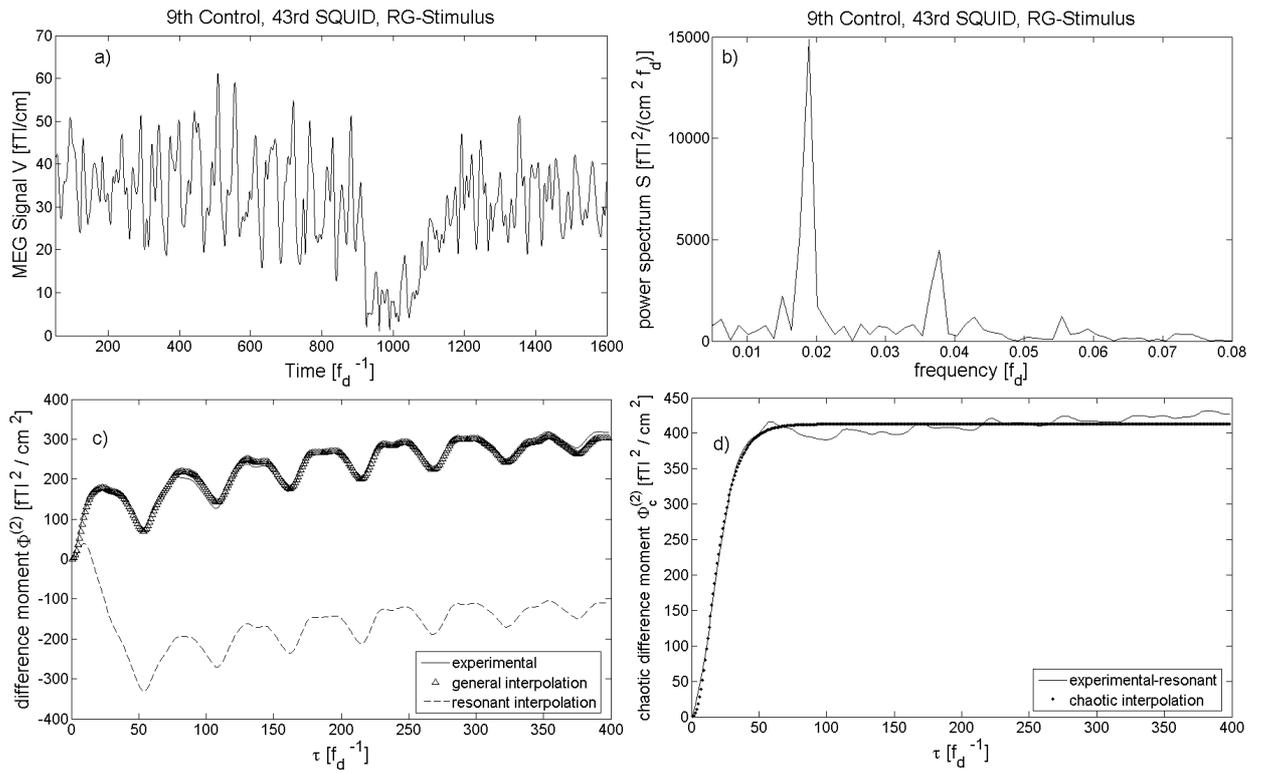

Fig. 14. Analysis of the MEG signal recorded at sensor 43 for control subject 9 as the response to RG-stimulus ($T = 3.2$ s; $\sigma = 14.4$ fTl/cm, $H_1 = 1.44$, $T_1 = 2.04 \cdot 10^{-2}$ s, $D \approx 1.02 \cdot 10^{4}$ fTl$^2$/(cm$^2$ s), $S_{cS}(0) = 1.34 \cdot 10^4$ fTl$^2$ / (cm$^2$ $f_d$), $T_0 = 2.42 \cdot 10^{-2}$ s, $n_0 = 4$): (a) source signal; (b) power spectrum $S(f)$ given by Eq. (3) in the low-frequency range (main peaks: 10 – 19 Hz); (c) "experimental" – $\Phi^{(2)}(\tau)$ given by Eq. (4), "general interpolation" – $\Phi^{(2)}(\tau)$ given by Eq. (18), "resonant interpolation" – $\tilde{\Phi}_r^{(2)}(\tau)$ given by Eq. (17); (d) – "experimental – resonant" – $\Phi^{(2)}(\tau)$ given by Eq. (18) minus $\tilde{\Phi}_r^{(2)}(\tau)$ given by Eq. (17), "chaotic interpolation" – $\Phi_c^{(2)}(\tau)$ given by Eq. (6).

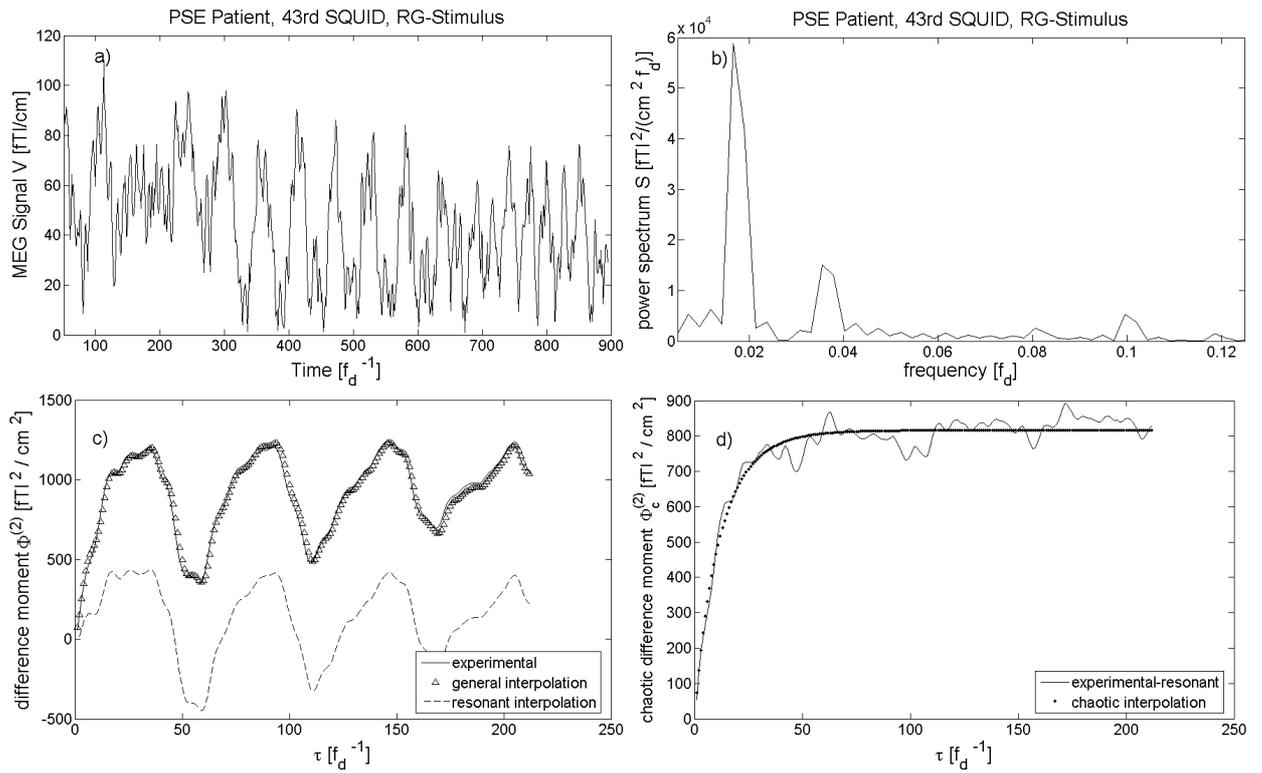

Fig. 15. Analysis of the MEG signal recorded at sensor 43 for the PSE patient as the response to RG-stimulus ($T$ = 1.7 s; $\sigma$ = 20.2 fTl/cm, $H_1$ = 0.47, $T_1$ = 3.2·10$^{-2}$ s, $D \approx$ 1.27·10$^4$ fTl$^2$/(cm$^2$ s), $S_{cS}(0)$ = 1.73·10$^4$ fTl$^2$ / (cm$^2 f_d$), $T_0$ = 1.97·10$^{-2}$ s, $n_0$ = 2.2): (a) source signal; (b) power spectrum $S(f)$ given by Eq. (3) in the low-frequency range (main peaks: 2.5 – 6 – 17.9 – 50 Hz); (c) "experimental" – $\Phi^{(2)}(\tau)$ given by Eq. (4), "general interpolation" – $\Phi^{(2)}(\tau)$ given by Eq. (18), "resonant interpolation" – $\tilde{\Phi}_r^{(2)}(\tau)$ given by Eq. (17); (d) – "experimental – resonant" – $\Phi^{(2)}(\tau)$ given by Eq. (18) minus $\tilde{\Phi}_r^{(2)}(\tau)$ given by Eq. (17), "chaotic interpolation" – $\Phi_c^{(2)}(\tau)$ given by Eq. (6).